\documentclass{article}

\usepackage{arxiv}

\usepackage{amsmath, amssymb} 
\usepackage{hyperref} 
\usepackage{array} 
\usepackage{multirow}
\usepackage{caption}

\usepackage{graphicx}

\usepackage[authoryear,comma,longnamesfirst,sectionbib]{natbib} 

\usepackage[flushleft]{threeparttable} 
\usepackage{xcolor, colortbl} 

\newcommand\independent{\protect\mathpalette{\protect\independenT}{\perp}}
\def\independenT#1#2{\mathrel{\rlap{$#1#2$}\mkern2mu{#1#2}}}
\newcolumntype{C}[1]{>{\centering\let\newline\\\arraybackslash\hspace{0pt}}m{#1}}

\title{Modeling Variables with a Detection Limit using a Truncated Normal Distribution with Censoring}

\author{
	Justin R.~Williams\\
	Department of Biostatistics\\
	University of California, Los Angeles\\
\And
	Hyung-Woo~Kim\\
	Solid Biosciences, Inc.\\
	Cambridge, Massachusetts\\
\And
	Catherine M.~Crespi\\
	Department of Biostatistics\\
	University of California, Los Angeles}

\renewcommand{\headeright}{}

\begin{document}

\maketitle

\begin{abstract}
When data are collected subject to a detection limit, observations below the detection limit may be considered censored. In addition, the domain of such observations may be restricted; for example, values may be required to be non-negative. We propose a regression method for censored observations that also accounts for domain restriction. The method finds maximum likelihood estimates assuming an underlying truncated normal distribution. We show that our method, tcensReg, outperforms other methods commonly used for data with detection limits such as Tobit regression and single imputation of the detection limit or half detection limit with respect to bias and mean squared error under a range of simulation settings. We apply our method to analyze vision quality data collected from ophthalmology clinical trials comparing different types of intraocular lenses implanted during cataract surgery.
\end{abstract}


\section{Introduction}\label{intro}
Censoring, in which the value of an observation is not known exactly but rather is only known to be above or below a specific value, is prevalent in many data settings. Censoring occurs with time-to-event data, but can also occur when measurements are subject to a detection limit (DL). A detection limit is defined as the lowest quantity or concentration of a compound that can be reliably detected with a given analytical method \citep{hornung1990estimation}. Quantities below the DL can be considered censored. Detection limits and the censored observations associated with them are encountered in epidemiology \citep{lubin2004epidemiologic, schisterman2006limitations}, hydrology \citep{helsel1990less}, chemistry \citep{analytical1987recommendations}, toxicology \citep{zaugg_usgs}, and economics \citep{mcdonald1980uses, greene_econometrics}.  

Estimation of the parameters of a normal distribution based on a sample with censored observations has a long history of investigation. \cite{hald1949maximum} was one of the first authors to develop maximum likelihood estimation methods for this data setting. Much of the early work focused on single mean models \citep{hald1949maximum, gupta1952estimation, harter1966iterative, tiku1967estimating} or estimation using order statistics \citep{sarhan1956estimation, dixon1960simplified}. Regression with a dependent variable subject to censoring gained prominence with the work of \cite{tobin1958estimation}, who developed the Tobit model, also called a censored regression or Tobit regression model, to estimate linear relationships between variables when there is censoring in the dependent variable. Linear regression methods with an unspecified censored distribution were developed by \cite{buckley1979linear}. 

In settings in which censored data arise due to a DL, estimation is sometimes performed by singly imputing the DL or 1/2 DL for observations below the DL. While these methods are known to yield biased estimates of the mean and standard deviation, they are still routinely applied due to convenience. Single imputation of the DL yields upwardly biased estimates of the mean \citep{helsel1990less}. There can also be substantial bias using 1/2 DL imputed values, with the direction of the bias depending on the underlying data mechanism \citep{hornung1990estimation, helsel1990less, lubin2004epidemiologic}. \cite{lubin2004epidemiologic} noted that the bias of parameter estimates when using 1/2 DL is substantial unless the proportion of censored observations is small, defined as less than 10\%. 

Related to but distinct from censoring is the concept of truncation. A truncated distribution is a conditional distribution that results from restricting the domain of some other probability distribution. For example, the zero-truncated Poisson distribution is the distribution of a Poisson random variable conditional on the value of the random variable being non-zero.  Truncation is a strict restriction of the domain of the random variable; observations outside the domain cannot occur.  In contrast, when an observation is censored, its true value is known to lie beyond the censoring threshold, and such true values are permitted to occur. Truncation is a property of the population, while censoring is a result of the sampling mechanism.

Methods have been developed for estimating the parameters of a truncated normal distribution. \cite{hald1949maximum} developed methods for the single mean model, while later work by \cite{cohen1950estimating} provided solutions for settings involving normal distributions with known or unknown truncation points and single or double truncation. \cite{halperin1952maximum} extended the theoretical framework to settings in which the distribution satisfies regularity conditions but is not necessarily normal. Work by \cite{amemiya1973regression} expanded the truncated regression framework by demonstrating the consistency and asymptotic normality properties of the maximum likelihood estimator and identifying consistent initial estimators.

Both censored and truncated data are often referred to as limited dependent variables in the economics literature \citep{tobin1958estimation, greene_econometrics}. There are currently methods of parameter estimation applicable to normal data that are censored and to data that are from a truncated normal distribution. However, estimation for the case in which both truncation and censoring are simultaneously generating the data, i.e. an underlying truncated normal distribution that is censored, are lacking. In this manuscript we propose maximum likelihood estimation techniques for such data. The methods can be seen as analogous to those of \cite{tobin1958estimation} with the latent normal distribution replaced with a latent truncated normal distribution. 

The paper is organized as follows. In Section~\ref{example}, a motivating example from an analysis of visual quality data from clinical trials of intraocular lenses implanted during cataract surgery is introduced. In Section~\ref{methods} the methods for maximum likelihood estimation are outlined. Subsection~\ref{single_mean} motivates the problem from a distributional perspective by defining the assumptions and the mechanism for censoring and truncation and develops the framework for estimation of a single mean model. In Subsection~\ref{lin_tcensReg}, methodology is developed for linear regression with left censored data from a left truncated normal distribution. In Subsection~\ref{tcensReg_sepvar}, the regression framework is extended to allow for heteroskedastic variances. In Subsection~\ref{opt}, methods to iteratively solve the log-likelihood for the parameters of interest are discussed. In Section~\ref{simulation}, the proposed methodology is compared to other methods in a simulation study. Section~\ref{application} applies the methods to our motivating example. Finally, Section~\ref{discussion} reflects on the significance of the results, discusses limitations and notes possible areas for extending the work.

\section{Problem Motivation}\label{example}
Our application concerns left censored non-negative data arising from contrast sensitivity testing in clinical trials for intraocular lenses implanted during cataract surgeries.

Contrast sensitivity measures the visual quality experienced by a subject by testing his or her ability to distinguish increasingly finer increments of light versus dark. Being unable to distinguish objects when contrast is low, i.e., when there is little difference between light and dark, can make everyday tasks such as night driving, navigating new settings, or perceiving distances difficult \citep{owsley1987contrast}. Cataract patients tend to have especially poor contrast sensitivity due to the clouding of the natural lens. The primary treatment for cataracts is cataract surgery, during which the natural clouded lens is removed and replaced with a new synthetical intraocular lens (IOL). The patient should see improvement in visual acuity and visual quality following IOL implantation. As such, contrast sensitivity is an important clinical outcome for patients who receive IOLs during cataract surgery.

Contrast sensitivity testing is performed using standardized charts with alternating light and dark bars, referred to as gratings. To determine the contrast sensitivity of a patient, the intensity of the contrast between the bars as well as the spacing is reduced until the patient is no longer able to perceive separate bars. The contrast is defined as the relative difference in luminance of the bars from the background and may be calculated using the Weber contrast, $\frac{L_{max}-L_{min}}{L_{background}}$, Michelson contrast, $\frac{L_{max}-L_{min}}{L_{max}+L_{min}}$, or RMS contrast, $\frac{L_{\sigma}}{L_{\mu}}$, where $L_{max}$, $L_{min}$, $L_{background}$, $L_{\mu}$, and $L_{\sigma}$ are luminance maximum, minimum, background, mean and standard deviation respectively \citep{pelli2013measuring}. Typically when using gratings to test contrast, the Michelson contrast is preferred. Contrast sensitivity is defined as the reciprocal of the threshold contrast, which is the lowest contrast to identify the grating. The spacing between the bars is measured in cycles per degree (CPD) with higher values of CPD indicating less space between bars. The testing is performed across a variety of different CPD levels, under either bright or dim lighting conditions, and with or without glare. In general, contrast sensitivity scores are lower when testing is performed under dim lighting with glare. 

\begin{figure}[ht]
\centering
\caption{CSV-1000E$^1$ Contrast Sensitivity Chart for 12.0 CPD}
\includegraphics[width = \textwidth, keepaspectratio]{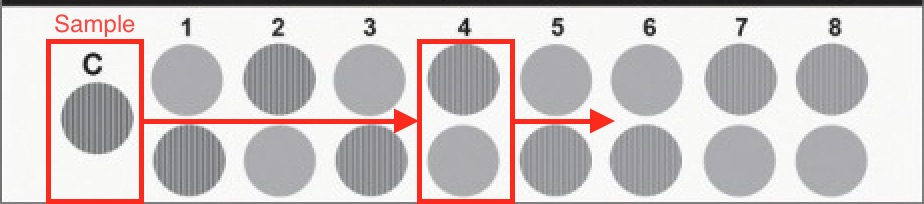}
\vspace*{2pt}
\caption*{$^1$ This testing chart is distributed by Vector Vision and was accessed from \url{http://www.vectorvision.com/csv1000-contrast-sensitivity/} on 29NOV2018}
\label{fig:twelve_cpd}
\end{figure}

At each CPD level, the subject is presented with a sample grating followed by 8 gratings that progressively decrease the intensity of the image contrast. Figure~\ref{fig:twelve_cpd} shows the testing setup used for 12.0 CPD, which is one of the most common visual quality outcomes analyzed in ophthalmic clinical trials.. The subject is first asked if they can identify the striped pattern in the sample image, and is subsequently shown each column starting from 1 and asked to identify whether the striped pattern is in the top, bottom, or neither grating. A contrast sensitivity score is recorded as the lowest level of identifiable contrast, i.e., the lowest intensity contrast for which the patient is able to correctly identify the striped pattern. The contrast sensitivity score ranges from 0 to 8, with 0 representing the ability to identify only the sample grating and 8 corresponding to the last column with the lowest image contrast. If a subject is unable to identify the sample grating, the value is recorded as -1. The scores of 0-8 are converted to continuous log contrast sensitivity values based on the manufacturers' recommendations as shown in Table~\ref{Tab:vv_logval}. Note that by definition, the contrast sensitivity values using the Michelson formula range from $[1,\infty)$, meaning that the log contrast sensitivity scores must be non-negative, i.e., $[0, \infty)$.

Problems arise using this scoring approach when patients are unable to identify the sample grating, i.e., have contrast sensitivity threshold score of -1. The true contrast sensitivity values for these patients are only known to be below the sample threshold. This is equivalent to having left censored observations. However, we also know that contrast sensitivity values must be non-negative, implying that the population values are left truncated at zero. Therefore, to accurately estimate the mean and standard deviation of contrast sensitivity for a particular IOL, both the censoring and truncation need to be accounted for.

Our data come from two prospective clinical trials for IOLs implanted during cataract surgery. The first study, \href{https://clinicaltrials.gov}{ClinicalTrials.gov} Identifier NCT01510717,  compared a monofocal lens and a multifocal lens in a double blind randomized parallel group study with bilateral IOL implantation, and the second study, \href{https://clinicaltrials.gov}{ClinicalTrials.gov} Identifier NCT01424189, compared two different multifocal lenses in a nonrandomized parallel assignment multi-center study again with bilateral IOL implantation. Our analysis will be restricted to data reported for binocular (both eyes open) testing, under dim lighting with and without glare. All observations were taken 6 months after surgery. 

\begin{figure}[ht]
\centering
\caption{Marginal Histograms for Monofocal and Multifocal Lens at 12 CPD under Dim Lighting}
\includegraphics[width = \textwidth, keepaspectratio]{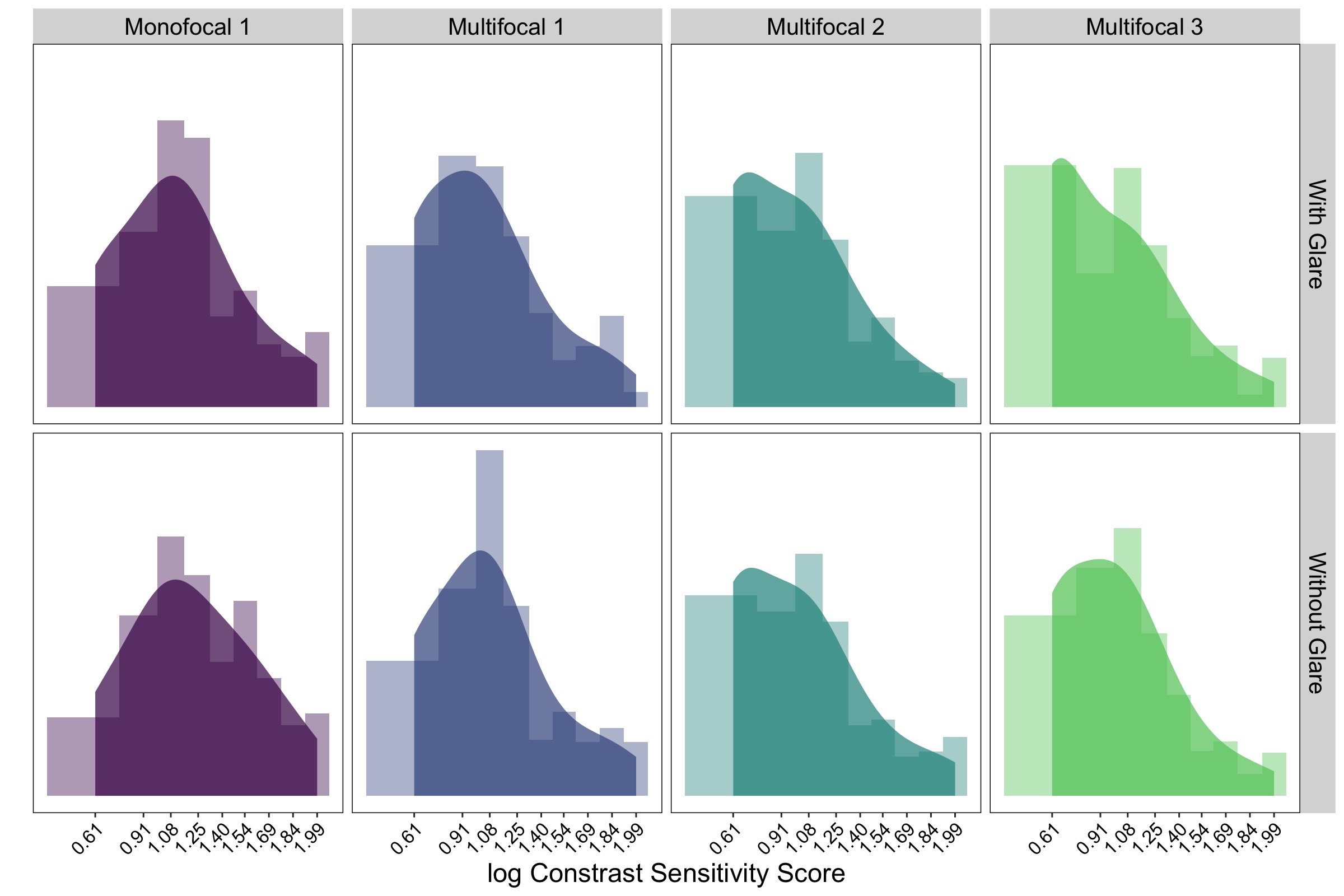}
\label{fig:marg_hist}
\caption*{Log contrast sensitivity scores are converted from contrast sensitivity threshold scores via Table~\ref{Tab:vv_logval}. Detection limit for 12 CPD occurs at $\nu=0.61$. Histogram is shown in the background with Gaussian kernel density estimate in the foreground with bandwidth set to 0.2.}
\end{figure}

Figure~\ref{fig:marg_hist} shows the marginal histograms with kernel density smoothers of each IOL by glare condition. Visual inspection of the data suggest that it may be reasonable to assume an underlying normal distribution for the log contrast sensitivity scores, with scores being censored at the lower-bound detection limit.

The goal of the analysis was to estimate the difference in mean log contrast sensitivity between the monofocal and each multifocal lenses. Historically, monofocal lenses have provided patients with better contrast sensitivity, but multifocal lenses are often preferred by patients for visual acuity as they provide both distance and near vision with increased spectacle independence. A common clinical trial hypothesis is to test whether the multifocal lens is non-inferior to the monofocal lens with respect to contrast sensitivity. Based on regulatory guidelines, losses of 0.3 log units are considered to be clinically significant when they occur at two or more spatial frequencies \citep{iso_tech_report}. The non-inferiority margin for contrast sensitivity is set at half of this clinically significant magnitude, i.e., loss of 0.15 log units. Our goal was thus to estimate pairwise differences in mean log contrast sensitivity between the monofocal lens and each multifocal lens to test for non-inferiority of visual quality.

\section{Methods}\label{methods}
This section first develops maximum likelihood estimation for a sample from a truncated normal distribution with censored observations, i.e., a single mean model. The methods are then extended to include a linear predictor for the mean to allow for linear regression applications. We then propose an extension to handle heteroskedastic variances. We focus on the setting in which the truncation and censoring occur only on the lower end of the distribution, i.e., left truncation and left censoring. The results can be generalized to right truncated and right censoring. Finally, the process of finding the maximum likelihood estimates is described using different optimization techniques.

\subsection{Single Mean Model}\label{single_mean}
As a first step, we obtain the likelihood function for censored data from a normal distribution. Assume a latent normal distribution with mean $\mu$ and variance $\sigma^{2}$ for the random variable $X^{*}$.
Following \cite{tobin1958estimation}, assume that the values of $X^{*}$ are left-censored at $\nu$, $\nu\in\mathbb{R}$, to produce the random variable $X$ defined as 
\begin{equation}\label{eqn:tobin_cens}
X_{i}=\left\{\begin{array}{lr}
\nu  &\text{if} \ X_{i}^{*}\le\nu,\\
X_{i}^{*} &\text{if} \ X_{i}^{*}>\nu.
\end{array}\right.
\end{equation}
Here, censored observations are reported as the DL $\nu$. The values of $X$ represent the observed values in the sample, which are a partial representation of the values of $X^{*}$. Assume that a total of $n$ observations are independently drawn with $n_0$ observations censored, i.e., $n_0=\sum_{i=1}^{n}1_{\{x_{i}=\nu\}}$, and $n_1$ observations uncensored, i.e., $n_1=\sum_{i=1}^{n}1_{\{x_{i}>\nu\}}$. The likelihood function for such data is
\begin{equation*}
L(\mu,\sigma)=\bigg[\Phi\Big(\frac{\nu-\mu}{\sigma}\Big)\bigg]^{n_0}\bigg[\frac{1}{\sigma}\bigg]^{n_1}\prod_{i\in S_{1}}\phi\Big(\frac{x_{i}-\mu}{\sigma}\Big),
\end{equation*}
where $\phi(\cdot)$ and $\Phi(\cdot)$ denote the standard normal pdf and cdf and $S_{1}$ is the set of all uncensored observations. Therefore the log-likelihood is
\begin{equation*}
l(\mu,\sigma)= n_0\ln\bigg[\Phi\Big(\frac{\nu-\mu}{\sigma}\Big)\bigg]-n_1\ln(\sigma)+\sum_{i\in S_{1}}\ln\bigg[\phi\Big(\frac{x_{i}-\mu}{\sigma}\Big)\bigg].
\end{equation*}
Maximum likelihood estimates $\hat{\mu}$ and $\hat{\sigma}$ can be found using iterative optimization techniques, such as the Newton Raphson algorithm, as discussed in \cite{tobin1958estimation}.

Now we replace the latent normal distribution above with a latent truncated normal distribution, truncated from below at the value $a$. Call this random variable $Y^{*}$, which has the following pdf and cdf from \cite{cohen1950estimating}:
\begin{equation*}
\begin{split}
f_{Y_{i}^{*}}(y_{i}^{*})=&\frac{1}{1-\Phi(\frac{a-\mu}{\sigma})}\bigg[\frac{1}{\sigma}\phi\Big(\frac{y_{i}^{*}-\mu}{\sigma}\Big)\bigg],\\
F_{Y_{i}^{*}}(y_{i}^{*})=&\frac{\Phi(\frac{y_{i}^{*}-\mu}{\sigma})-\Phi(\frac{a-\mu}{\sigma})}{1-\Phi(\frac{a-\mu}{\sigma})}.
\end{split}
\end{equation*}
The distribution of $Y^{*}$ is a scaled version of a normally distributed random variable, obtained by dividing the pdf by the constant $1-\Phi(\frac{a-\mu}{\sigma})$ to obtain a proper probability density function that integrates to one. We will denote a latent truncated normal random variable with left truncation at a constant $a\in\mathbb{R}$ as
\begin{equation*}
Y^{*}\sim\text{TN}(\mu, \sigma^{2}, a),
\end{equation*}
where we assume that the truncation value $a$ is known and therefore fixed. For non-negative variables, $a=0$. 

Note that the parameter $\mu$ denotes the mean of the underlying normal distribution prior to truncation, rather than the mean of the truncated normal distribution, which is $\mu_{TN}=\mu+\frac{\phi(\frac{a-\mu}{\sigma})}{1-\Phi(\frac{a-\mu}{\sigma})}\sigma$ \citep{greene_econometrics}. Throughout the paper we focus on the estimation of this underlying central tendency parameter $\mu$ rather than $\mu_{TN}$.

The log-likelihood for the truncated normal distribution with $n$ independent observations drawn from this distribution is
\begin{equation*}
l(\mu,\sigma)=-n\ln\bigg[1-\Phi\Big(\frac{a-\mu}{\sigma}\Big)\bigg]-n\ln(\sigma)+\sum_{i=1}^{n}\ln\bigg[\phi\Big(\frac{y_{i}^{*}-\mu}{\sigma}\Big)\bigg].
\end{equation*}
There is no closed form for the maximum likelihood estimates for $\mu$ and $\sigma$ for the truncated normal distribution. However, \cite{cohen1950estimating} described methods for iteratively solving for the maximum likelihood estimates and \cite{amemiya1973regression} showed that the estimates were consistent and asymptotically normal.

Further suppose that the values of $Y^{*}$ are censored at $\nu>a$, $\nu\in\mathbb{R}$, to produce the random variable $Y$, defined as
\begin{equation} \label{eqn:obs_cens}
Y_{i}=\left\{\begin{array}{lr}
\nu  &\text{if} \ Y_{i}^{*}\le\nu,\\
Y_{i}^{*} &\text{if} \ Y_{i}^{*}>\nu,
\end{array}\right.
\end{equation}
where $\nu$ is a known constant. For example in the contrast sensitivity problem, the testing procedure has an inherent detection limit and cannot detect values below $0.61$ log units. The log contrast sensitivity scores must also be non-negative as discussed earlier in Section~\ref{example}.

\begin{figure}[ht]
\centering
\caption{Truncated Normal Distribution with Censoring}
\includegraphics[width = \textwidth, keepaspectratio]{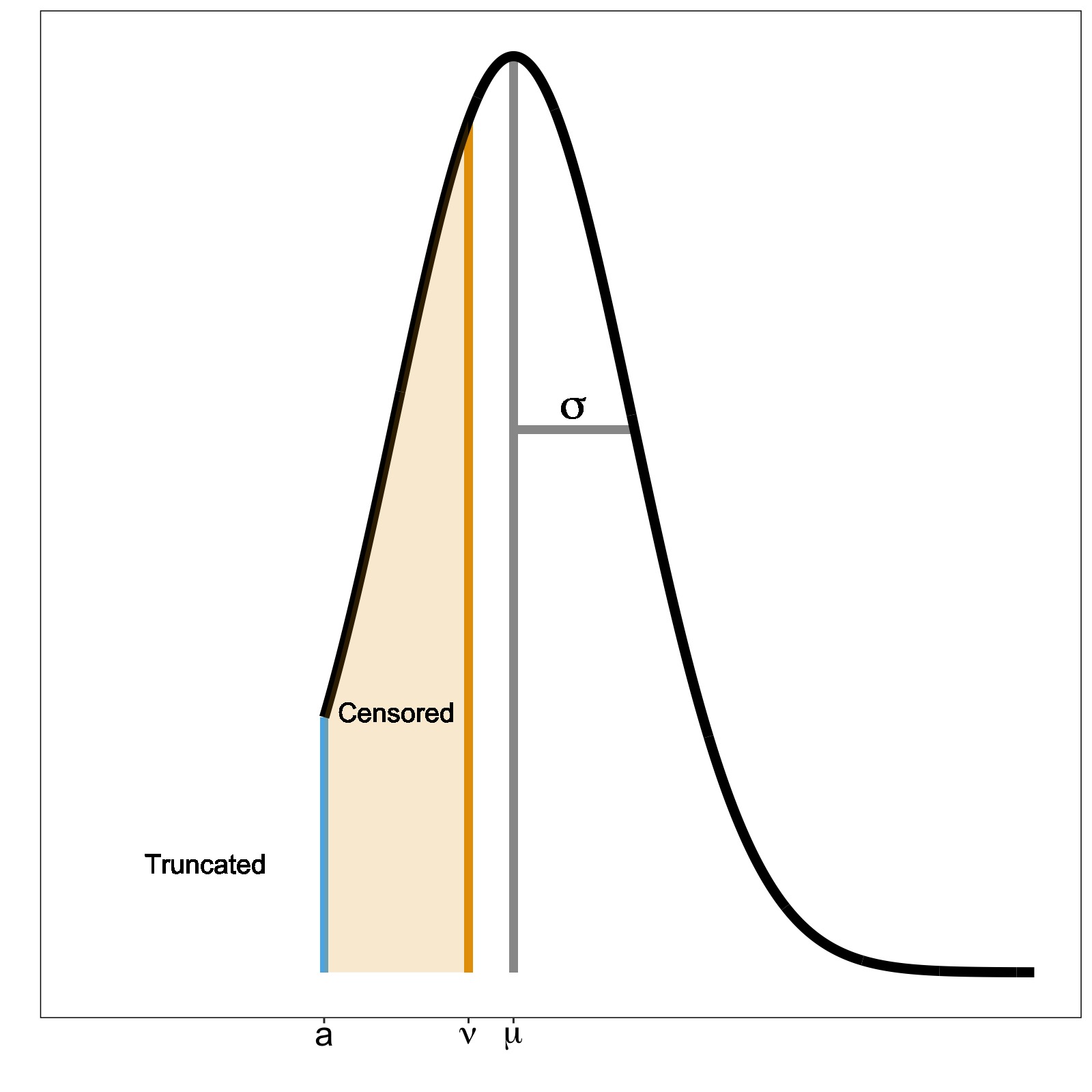}
\label{fig:trunc_cens}
\caption*{Potential density for a left truncated normal distribution with left censoring. The density above was created with $\mu=0.8$, $\sigma=0.5$, $a=0$, and $\nu=0.61$. In our application, $a=0$ and $\nu=0.61$.}
\end{figure}

The pdf for the truncated random variable with censoring can be expressed as
\begin{equation}
f_{Y_{i}}(y_{i})=1_{\{y_{i}=\nu\}}\Bigg[\frac{\Phi\Big(\frac{\nu-\mu}{\sigma}\Big)-\Phi\Big(\frac{a-\mu}{\sigma}\Big)}{1-\Phi(\frac{a-\mu}{\sigma})}\Bigg]+1_{\{y_{i}>\nu\}}\Bigg[\frac{1}{\sigma\Big(1-\Phi(\frac{a-\mu}{\sigma})\Big)}\phi\Big(\frac{y_{i}-\mu}{\sigma}\Big)\Bigg].
\end{equation}
The first term of this equation captures censored observations, in which case the data are reported as the detection limit value $\nu$ and the cdf of the truncated normal is used to provide information for the likelihood. In the second term, the observation is not censored and we simply use the pdf for the truncated normal. Figure~\ref{fig:trunc_cens} shows an example of the probability density function for such a random variable.

Out of a total of $n$ observations, again let $n_0$ be censored such that $n_0=\sum_{i=1}^{n}1_{\{y_{i}=\nu\}}$, and $n_1$ be uncensored, $n_1=\sum_{i}^{n}1_{\{y_{i}>\nu\}}$. Let $S$ define the set of all observations, $S_{1}$ be the set of uncensored observations, and $S_{0}$ be the set of censored observations, i.e., $S_{0}\cup S_{1}=S$. Assuming that the observations are drawn independently, the likelihood is
\begin{equation}
L(\mu,\sigma)=\Bigg(\frac{1}{1-\Phi(\frac{a-\mu}{\sigma})}\Bigg)^{n}\bigg[\Phi\Big(\frac{\nu-\mu}{\sigma}\Big)-\Phi\Big(\frac{a-\mu}{\sigma}\Big)\bigg]^{n_0}\Big(\frac{1}{\sigma}\Big)^{n_1}\prod_{i\in S_{1}}\phi\Big(\frac{y_{i}-\mu}{\sigma}\Big).
\end{equation}

Taking the log of the likelihood, we have
\begin{equation}\label{eqn:log_like}
\begin{split}
l(\mu,\sigma)=&-n\ln\bigg[1-\Phi\Big(\frac{a-\mu}{\sigma}\Big)\bigg] +n_0\ln\bigg[\Phi\Big(\frac{\nu-\mu}{\sigma}\Big)-\Phi\Big(\frac{a-\mu}{\sigma}\Big)\bigg]-n_1\ln(\sigma)\\
&+\sum_{i\in S_{1}}\ln\bigg[\phi\Big(\frac{y_{i}-\mu}{\sigma}\Big)\bigg].
\end{split}
\end{equation}

Similar to the censored only and truncated only log-likelihoods, the maximum likelihood estimators for $\mu$ and $\sigma$ for the truncated normal distribution with censoring do not have a closed form but can be estimated using the iterative process discussed in Section~\ref{opt}.

\subsection{Linear Predictor for the Mean}\label{lin_tcensReg}
The goal of many applications is to understand how certain predictors influence the mean or to compare the means of different populations. This can be accomplished by using the linear predictor $\mathbf{X}_{i}^{T}\boldsymbol{\beta}$, i.e.,
\begin{equation*}
Y_{i}^{*}\overset{iid}{\sim}\text{TN}(\mathbf{X}_{i}^{T}\boldsymbol{\beta}, \sigma^{2}, a),
\end{equation*}
where $\boldsymbol{\beta}=(\beta_{1},\dots,\beta_{p-1})^{T}$ is a $(p-1)\times 1$ vector of parameters with $p\ge 2$. Again, note that $\mathbf{X}_{i}^{T}\boldsymbol{\beta}$ is the mean of the underlying normal distribution rather than the mean of the truncated normal distribution, that is, $\text{E}[X_{i}^{*}]$ rather than $\text{E}[Y_{i}^{*}]$. All of the unknown parameters can be collected into the vector $\boldsymbol{\boldsymbol{\theta}} = (\boldsymbol{\beta},\sigma^{2})^{T}$ which has length $p$. The corresponding pdf and cdf are
\begin{equation*}
\begin{split}
f_{Y_{i}^{*}}(y_{i}^{*})=&\frac{1}{1-\Phi(\frac{a-\mathbf{x}_{i}^{T}\boldsymbol{\beta}}{\sigma})}\bigg[\frac{1}{\sigma}\phi\Big(\frac{y_{i}^{*}-\mathbf{x}_{i}^{T}\boldsymbol{\beta}}{\sigma}\Big)\bigg]\\
F_{Y_{i}^{*}}(y_{i}^{*})=&\frac{\Phi(\frac{y_{i}^{*}-\mathbf{x}_{i}^{T}\boldsymbol{\beta}}{\sigma})-\Phi(\frac{a-\mathbf{x}_{i}^{T}\boldsymbol{\beta}}{\sigma})}{1-\Phi(\frac{a-\mathbf{x}_{i}^{T}\boldsymbol{\beta}}{\sigma})}.
\end{split}
\end{equation*}

Suppose that this truncated normal distribution is then censored at the value $\nu$, where $\nu > a$. Let the notation $a_{i}^{*}=\frac{a-\mathbf{x}_{i}^{T}\boldsymbol{\beta}}{\sigma}$ denote a standardized version of the constant $a$. The likelihood of the truncated normal distribution with censoring can be expressed as
\begin{equation}\label{eqn:log_like_lin}
\begin{split}
l(\boldsymbol{\beta},\sigma^{2})=&-\sum_{i=1}^{n}\ln\Big[1-\Phi(a_{i}^{*})\Big]+\sum_{i\in S_{0}}\ln\Big[\Phi(\nu_{i}^{*})-\Phi(a_{i}^{*})\Big] - n_{1}\ln(\sigma)\\
&+\sum_{i\in S_{1}}\ln\Big[\phi\Big(\frac{y_{i}-\mathbf{x}_{i}^{T}\boldsymbol{\beta}}{\sigma}\Big)\Big].
\end{split}
\end{equation}

\subsection{Heteroskedastic Variances}\label{tcensReg_sepvar}
We now relax the assumption of homogeneous variance. We consider the case of independent groups with different variances. Assume we have samples drawn independently from $J$ truncated normals, with each population having a common truncation value but possibly different variance, according to the model
\begin{equation*}
Y^{*}_{ij}\sim\text{TN}(\mathbf{X}_{ij}^{T}\boldsymbol{\beta}, \sigma^{2}_{j}, a), \quad i = 1,\dots,n_{j} \ \text{and} \ j = 1,\dots,J ,
\end{equation*}
where $n_{j}$ is the number of observations in group $j$ and $Y^{*}_{ij} \independent Y^{*}_{i'j'}$ for all $i\ne i'$ and $j\ne j'$. Assume that observations are censored at the value $\nu$ to create a sample of independent random variables with pdf
\begin{equation*}
f_{Y_{ij}}(y_{ij})=1_{\{y_{ij}=\nu\}}\Bigg[\frac{\Phi\Big(\frac{\nu - \mathbf{x}_{ij}^{T}\boldsymbol{\beta}}{\sigma_{j}}\Big)-\Phi\Big(\frac{a - \mathbf{x}_{ij}^{T}\boldsymbol{\beta}}{\sigma_{j}}\Big)}{1-\Phi\Big(\frac{a - \mathbf{x}_{ij}^{T}\boldsymbol{\beta}}{\sigma_{j}}\Big)}\Bigg]+1_{\{y_{ij}>\nu\}}\Bigg[\frac{1}{\sigma_{j}\Big(1-\Phi(\frac{a - \mathbf{x}_{ij}^{T}\boldsymbol{\beta}}{\sigma_{j}})\Big)}\phi\Big(\frac{y_{ij}-\mathbf{x}_{ij}^{T}\boldsymbol{\beta}}{\sigma_{j}}\Big)\Bigg].
\end{equation*}
Because groups are independent, the log likelihood becomes
\begin{equation}\label{eqn:log_like_sepvar}
\begin{split}
l(\boldsymbol{\beta},\sigma^{2}_{1},\dots,\sigma^{2}_{J})=&\sum_{j=1}^{J}\sum_{i=1}^{n_{j}}-\ln\Big[1-\Phi(a_{ij}^{*})\Big]+\sum_{i\in S_{0j}}\ln\Big[\Phi(\nu_{ij}^{*})-\Phi(a_{ij}^{*})\Big] - n_{1j}\ln(\sigma_{j})\\
&+\sum_{i\in S_{1j}}\ln\Big[\phi\Big(\frac{y_{ij}-\mathbf{x}_{ij}^{T}\boldsymbol{\beta}}{\sigma_{j}}\Big)\Big],
\end{split}
\end{equation}
where $a_{ij}^{*}=\frac{a-\mathbf{x}_{ij}^{T}\boldsymbol{\beta}}{\sigma_{j}}$, $S_{0j}$ and $S_{1j}$ are the sets of censored observations and uncensored observations respectively in the $j^{th}$ group, and $n_{1j}$ is the number of uncensored observations in the $j^{th}$ group. 

\subsection{Obtaining Maximum Likelihood Estimates}\label{opt}
Our goal is to find the values $\hat{\boldsymbol{\theta}}$ that maximize the log-likelihoods of Equations~\ref{eqn:log_like},~\ref{eqn:log_like_lin}, and~\ref{eqn:log_like_sepvar}. However, closed form solutions do not exist. To obtain maximum likelihood estimates, an iterative procedure is required. One approach is to use the Newton Raphson algorithm using Taylor series expansion, as discussed in Chapter 14 in \cite{lange2010numerical}. While the Newton-Raphson method has attractive local convergence guarantees and reliable performance \citep{ypma1995historical}, each step requires evaluation of the Hessian matrix, which can become computationally expensive for a large set of predictors. Alternative optimization routines, such as the quasi-Newton BFGS \citep{BFGS_algorithm_broyden, BFGS_algorithm_fletcher, BFGS_algorithm_goldfarb, BFGS_algorithm_shanno} or the conjugate gradient \citep{CG_algorithm}, which require only evaluation of the likelihood function and corresponding gradient, often require additional evaluations but have reduced memory and faster computing time. 

Within R, other optimization packages such as the maxLik package from \cite{henningsen2011maxlik} can also be used to find maximum likelihood estimates. This package is called in the censored only and truncated only maximum likelihood estimation packages in R such as censReg by \cite{henningsen2010estimating} and truncreg by \cite{croissant2018truncreg}. 

We developed the standalone R package \href{https://github.com/williazo/tcensReg}{tcensReg} to solve the novel likelihood equation of the truncated normal distribution with censoring. This software package uses analytic results of the gradient and Hessian for the corresponding model of interest in either Equation~\ref{eqn:log_like_lin} or Equation~\ref{eqn:log_like_sepvar} derived in Appendices~\ref{grad_simple},~\ref{hess_simple},~\ref{grad_sepvar}, and~\ref{hess_sepvar}. Several optimization routines are available within the software including conjugate gradient, Newton-Raphson, and BFGS. This package uses familiar model syntax and has additional functionality to estimate parameters for the censored only or truncated only settings similar to the censReg and truncreg packages, respectively.

\section{Simulation Study}\label{simulation}
We conducted a simulation study to compare the performance of our method to that of five methods of estimating the mean and standard deviation of the underlying normal distribution from a truncated normal distribution with censored observations. Method 1 is the gold standard method which uses the true uncensored observations, and accounts for the truncation in the estimation procedure by using the appropriate truncated log-likelihood. Method 2 uses the same uncensored truncated data but does not adjust for truncation in the normal distribution likelihood function. Methods 3-6 use the censored truncated observations but differ in how they treat censoring and truncation. Method 3 imputes all censored values with the detection limit and uses maximum likelihood estimation with a normal distribution likelihood, while Method 4 imputes all values as half of the detection limit and also uses normal maximum likelihood estimation. Both the DL and 1/2 DL imputation methods have been shown to perform poorly in cases with censoring \citep{hornung1990estimation, helsel1990less, lubin2004epidemiologic}, and are expected to show even worse performance in this setting since they do not account for censoring or truncation. However, they are still sometimes used in practice. Method 5 uses Tobit regression, which incorporates censoring into the likelihood. This method is often recommended when the assumption of normality seems reasonable. Finally, Method 6 is our proposed maximum likelihood estimation procedure described in Section~\ref{lin_tcensReg} which takes into account not only the censoring but also the truncation of the underlying distribution. These six methods will be referred to as: 1) Gold Standard (GS), 2) Uncensored with no truncation adjustment (Uncens NT), 3) Detection Limit (DL), 4) 1/2 Detection Limit (1/2 DL), 5) Tobit regression (Tobit), and 6) censored regression with truncation adjustment (tcensReg), our proposed method.

\subsection{Set-up}
Three different sets of simulation studies were conducted to compare the six methods. In the first simulation study, values from a single mean model were drawn to compare performance of the methods in terms of bias and mean squared error (MSE) for estimating the mean and standard deviation of the underlying latent distribution. The second simulation study focused on estimating the difference of the means of two independent populations and their common variance. Finally, the third simulation study assessed their performance in a non-inferiority test setting similar to that of our motivating example. The simulations were conducted in R version 3.5.1\citep{rcoreteam2018}.

For the first simulation study, values were simulated from a truncated normal distribution,
\begin{align*}
Y_{i}^{*}\sim\text{TN}(\mu, \sigma^{2}, a),
\end{align*}
and then censored. A constant value of $a = 0$ was used to represent zero-truncation. The values of $\mu$ and $\sigma^{2}$ were chosen to approximate the marginal distributions of the log contrast sensitivity scores from the application introduced in Section~\ref{example}. Typical values of mean log contrast sensitivity ranged from $0.7$ to $1.1$, and thus we used $\mu\in\{0.7, 0.8, 0.9, 1.0, 1.1\}$. Based on standard deviations observed in the contrast sensitivity data, for the simulation we used $\sigma\in\{0.40, 0.45, 0.50\}$. This created a total of $15$ $(5\times3)$ parameter combinations for $\mu$ and $\sigma^{2}$.

Observations from a truncated normal distribution, $y\sim\text{TN}(\mu,\sigma^{2}, a)$, can be simulated by transforming samples from a uniform distribution on the interval $[0, 1]$ using the inverse probability transformation:
\begin{equation*}
y=\Phi^{-1}\Bigg\{p\times\bigg[1-\Phi\big(\frac{a-\mu}{\sigma}\big)\bigg] + \Phi\big(\frac{a-\mu}{\sigma}\big)\Bigg\}\times\sigma+\mu,
\end{equation*}
where $p$ represents the sample from the uniform distribution \cite{burkardt2014truncated}. 
This method of inverse transform sampling is implemented with the tcensReg software package. After transforming to the appropriate truncated normal distribution, another dataset was generated by censoring the observations at $\nu=0.61$. Values that fell below $\nu$ were either replaced with the DL (0.61), 1/2 DL (0.305), or marked as censored for Tobit and tcensReg estimation.

Each of the six estimation methods was used to estimate $\mu$ and $\sigma$. The data simulation and estimation procedures were repeated for $B=10,000$ replications. For each of the six methods, two performance metrics were calculated for $\theta\in\{\mu, \sigma\}$: average bias ($\bar{\hat{\theta}}-\theta$) and mean squared error (MSE; $1/B\sum_{k=1}^{B}[\hat{\theta}_{k}-\theta]^{2}$) where $\bar{\hat{\theta}}=1/B\sum_{k=1}^{B}\hat{\theta}_{k}$.

In the second simulation study, we simulated values from two truncated normal populations with different means but common variance and truncation value, i.e.,
\begin{align*}
Y_{1}^{*}&\sim\text{TN}(\mu_{1}, \sigma^{2}, a) & Y_{2}^{*}&\sim\text{TN}(\mu_{2},\sigma^{2}, a).
\end{align*}
A constant value of $a = 0$ was used to represent zero-truncation and a range of values of $\mu_{1}$, $\mu_{2}$, and $\sigma^{2}$ were selected to produce data similar to the application. Population 1 approximated a monofocal intraocular lens, while Population 2 approximated a multifocal intraocular lens. Means for monofocal lenses took values $\mu_{1}\in\{1.0,1.1\}$, and the difference between multifocal and monofocal lenses was set to range from no difference to a clinically significant difference of 0.3 log units \citep{iso_tech_report}, i.e.,  $\delta\in\{-0.3, -0.2, -0.1, 0\}$. Therefore the mean of Population 2 was set as $\mu_{2}=\mu_{1} + \delta$. The common standard deviation was assumed to take values $\sigma\in\{0.40, 0.45, 0.50\}$. The two-population simulation had a total of $24$ $(2\times4\times3)$ parameter combinations for $\mu_{1}$, $\mu_{2},$ and $\sigma^{2}$. 

A total of 100 observations from each population were sampled. Again, for each population a separate censored dataset was generated using the censoring threshold $\nu=0.61$, and values that fell below $\nu$ were replaced with the DL ($ 0.61$), 1/2 DL ($0.305$) or marked as censored for Tobit and tcensReg estimation. Each of the six estimation methods was used to estimate the mean difference between Population 1 and 2, $\delta$, and the common standard deviation, $\sigma$. Data simulation and estimation were repeated for $B=10,000$ replications. For each of the six methods, average bias and MSE were calculated.

The third simulation study assessed the performance of the methods in the context of a non-inferiority test. Here we focused on the Type I error rates of the various methods. Type I error rates are particularly important for non-inferiority tests because, if non-inferiority is falsely accepted, patients may decide between products based on non-efficacy factors such as price and side effects assuming the products to be similar when in fact one is truly superior. For the non-inferiority test, data were simulated with a true difference of $\delta = -0.15$ in the two population model, while varying $\mu_{1} \in\{1.0, 1.1\}$ and $\sigma\in\{0.40, 0.45, 0.50\}$ for a total of 6 different non-inferiority scenarios. Then each of the six methods was used to construct $1-\alpha$ confidence intervals for $\delta$. The test is specified as one-sided because it is known that multifocal contrast sensitivity is less than monofocal contrast sensitivity. Constructing $1-\alpha$ confidence intervals and comparing the lower bound to the non-inferiority margin will result in a one-sided hypothesis test at the $\alpha$ level. If the lower bound of the confidence interval does not cover the true value of $\delta$, then the non-inferiority hypothesis would be falsely accepted. A total of 100 observations were drawn from each population to construct the confidence intervals. The hypothesis test was repeated for $B=10,000$ replications and the Type I error rate was calculated as the percent of replications where the lower bound of the $1-\alpha$ confidence interval was greater than $-0.15$.

The choice of initial starting values $\boldsymbol{\theta}^{(0)}$ is important since optimization algorithms can provide local rather than global convergence. To ensure that the starting values are reasonable, we recommend using initial estimates from a censored regression model. These estimates showed excellent rates of convergence for our simulation settings.

\subsection{Results}
Table~\ref{Tab:cens_pct_singlemean} shows expected censoring and truncation percentages and the ratio of censoring to truncation for each of the 15 parameter value scenarios in the single mean model. The expected percent of censoring ranges from 10.8\%-38.6\% and truncation was typically $\le5\%$ but was slightly higher when the mean was closer to the truncation value of $0$, i.e., $\mu\in\{0.7, 0.8\}$. The ratio of censoring to truncation varies from 4.68 to 35.87 where values greater than 1 indicate more censoring than truncation. In general, for a fixed $\mu$, as $\sigma$ increases the expected censoring and truncation increase while the ratio of censoring to truncation decreases. 

\begin{figure}[ht]
\centering
\caption{Average Bias for $\mu$ from Six Different Estimation Methods in Single Mean Model}
\includegraphics[width = \textwidth, keepaspectratio]{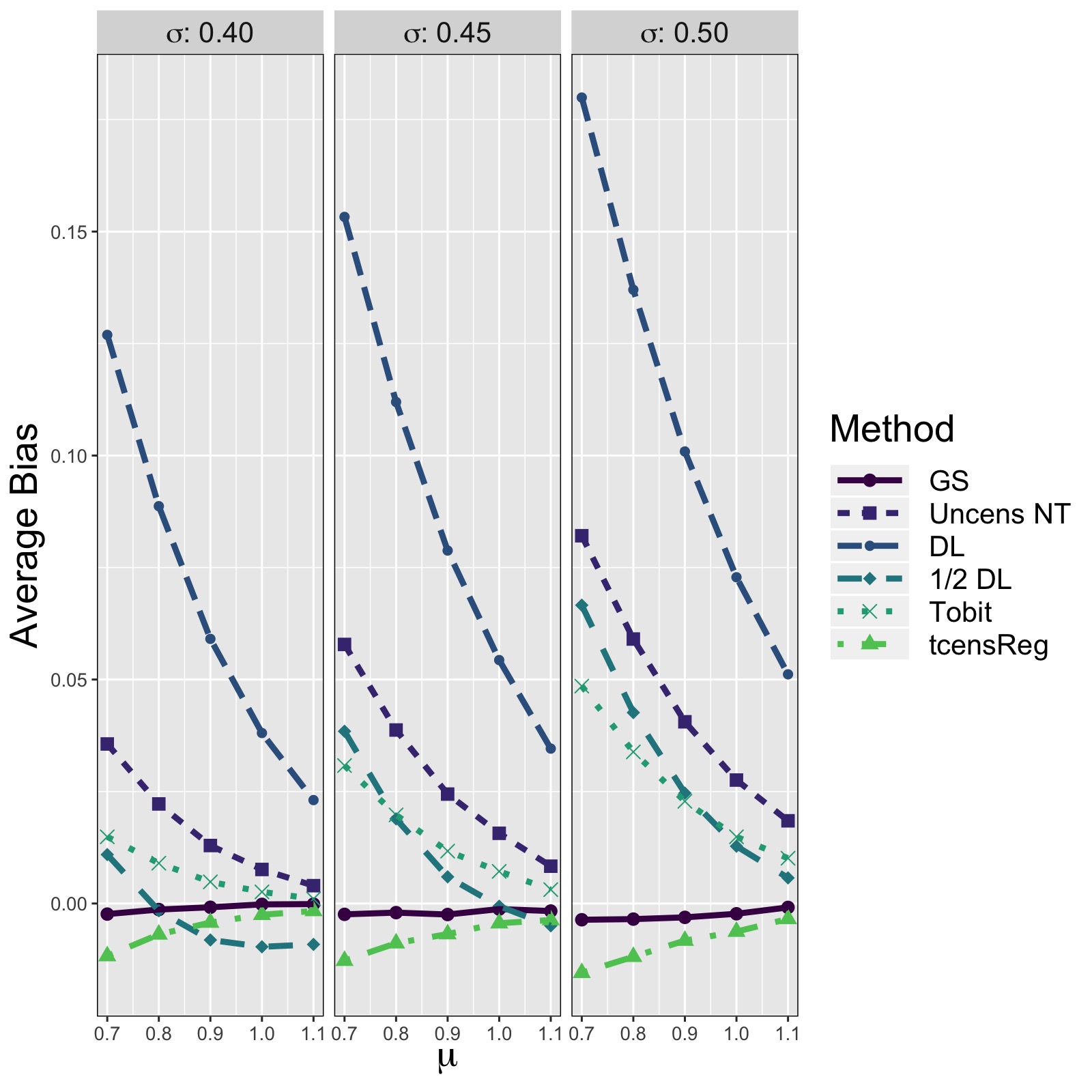}
\label{fig:tnorm_mu_bias}
\caption*{GS = Gold Standard, i.e. uncensored observations with truncation adjustment

Uncens NT = Uncensored data with no truncation adjustment

DL = detection limit

Tobit = Tobit censored regression with no truncation adjustment

tcensReg = Censored regression with truncation adjustment}
\end{figure}

Performance of the six methods in terms of average bias for $\mu$ is shown in Figure~\ref{fig:tnorm_mu_bias}. The figure shows that the gold standard was essentially unbiased with only slight negative bias for low values of $\mu$. The average bias of tcensReg was slightly negative for all scenarios, meaning that the estimated mean was smaller than the true value. The slight negative bias increased as the amount of censoring increased, i.e., as $\mu$ decreased for a fixed $\sigma$ and as $\sigma$ increased. However, the absolute bias remained small and was always below $2\%$ of the true $\mu$ value. The Tobit, Uncens NT and DL methods consistently had positive bias, with the estimated mean consistently larger than the true value. The amount of bias also increased more rapidly for these other methods as censoring increased, compared to tcensReg. The 1/2 DL method had variable trends in bias with positive bias in some settings and negative bias in others. In particular, when $\sigma=0.40$, the 1/2 DL method overestimated $\mu$ when $\mu\le0.8$ and underestimated it when $\mu>0.8$. 

The tcensReg method was consistently closer to the gold standard compared to the Tobit method, especially in situations with high censoring, i.e., $\sigma=0.50$. There were high censoring settings in which the tcensReg method and Tobit method performed similarly; for example, absolute bias was similar when $\mu=0.9$ and $\sigma=0.40$, which had 22.47\% expected censoring. For this scenario, the ratio of censoring to truncation is 18.42 meaning that there is more than 18 times more expected censoring than truncation. For cases where the ratio of censoring to truncation is low, the tcensReg method tends to outperform the Tobit method. When the ratio of censoring to truncation is greater, the Tobit method is comparable in terms of absolute bias to the tcensReg method, but with different directions of bias.

\begin{figure}[ht]
\centering
\caption{Average Log Mean Squared Error for $\mu$ from Six Different Estimation Methods in Single Mean Model}
\includegraphics[width = \textwidth, keepaspectratio]{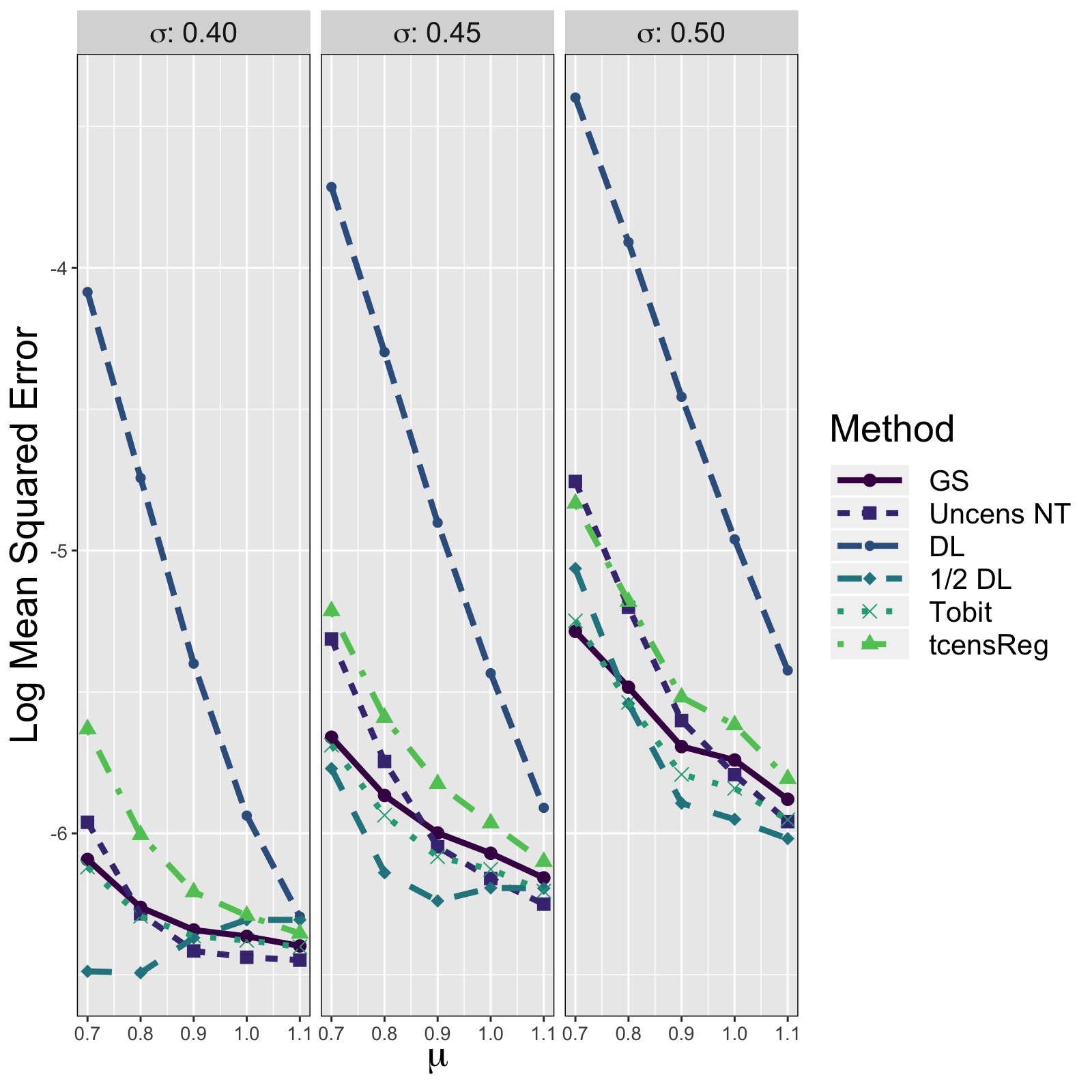}
\label{fig:tnorm_mu_mse}
\caption*{GS = Gold Standard, i.e. uncensored observations with truncation adjustment

Uncens NT = Uncensored data with no truncation adjustment

DL = detection limit

Tobit = Tobit censored regression with no truncation adjustment

tcensReg = Censored regression with truncation adjustment}
\end{figure}

The precision of the estimates of the mean as reflected by their log MSE is shown in Figure~\ref{fig:tnorm_mu_mse}. MSE was transformed to the log scale as there was a great disparity between the values for the DL point imputation method and the other five methods. Here, the average log MSE of each method should be compared to that of the gold standard. The DL method consistently had the highest log MSE of all of the methods. The tcensReg method had higher log MSE than the Tobit, 1/2 DL, and Uncens NT methods, suggesting that these methods were outperforming tcensReg. However, the average log MSE for these three methods often fell below that of the gold standard, which reflects an underestimation of the true sampling variance. In contrast, the tcensReg method always maintained a log MSE greater than or about equal to the gold standard. Overall, the tcensReg method avoided false precision.

\begin{figure}[ht]
\centering
\caption{Average Bias for $\sigma$ from Six Different Estimation Methods in Single Mean Model}
\includegraphics[width = \textwidth, keepaspectratio]{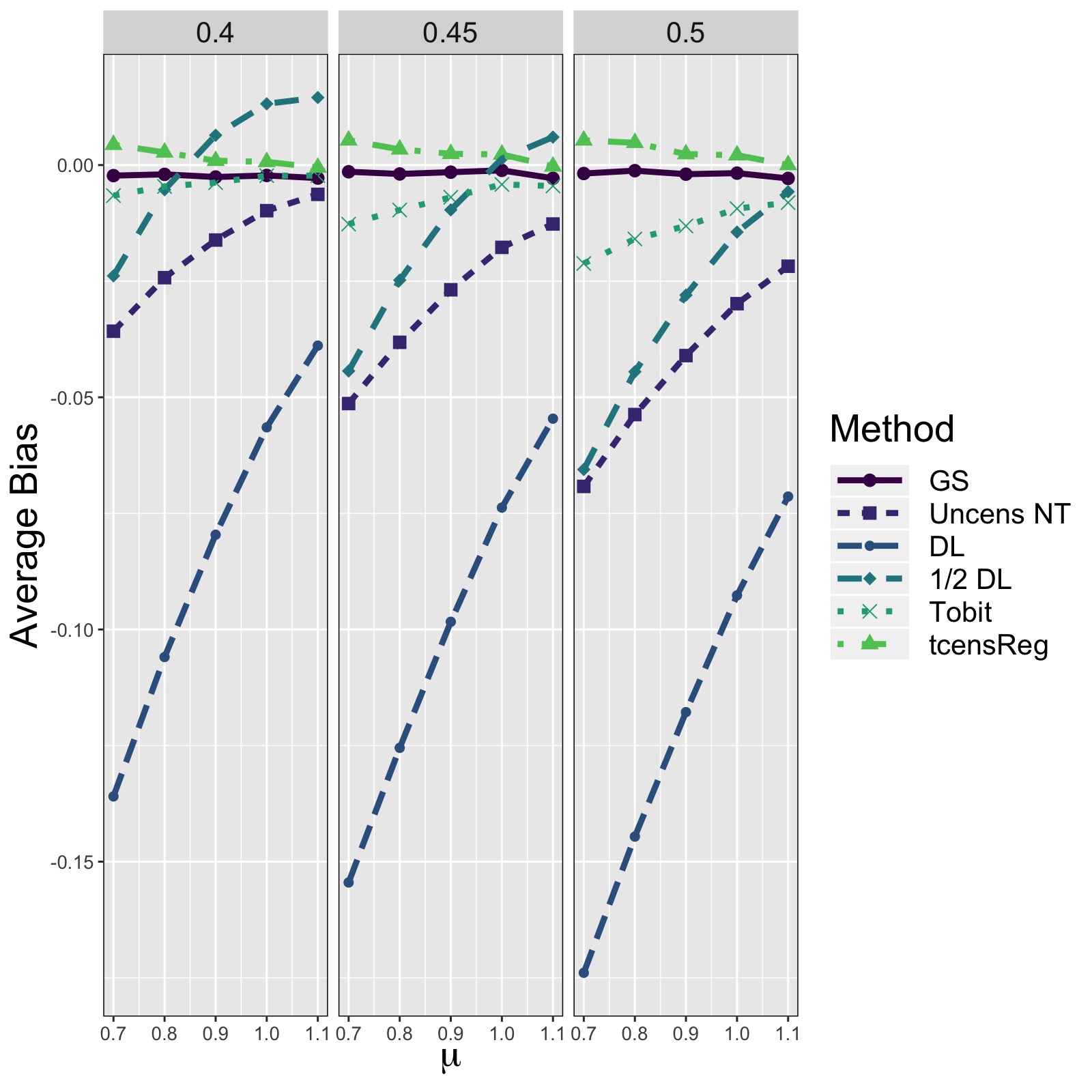}
\label{fig:tnorm_sd_bias}
\caption*{GS = Gold Standard, i.e. uncensored observations with truncation adjustment

Uncens NT = Uncensored data with no truncation adjustment

DL = detection limit

Tobit = Tobit censored regression with no truncation adjustment

tcensReg = Censored regression with truncation adjustment}
\end{figure}

Figure~\ref{fig:tnorm_sd_bias} provides evidence that other methods such as Tobit, 1/2 DL, and Uncens NT also systematically underestimated the variance of the underlying distribution. The Tobit and Uncens NT always have negative bias, meaning that these methods underestimated $\sigma$, with this bias being especially pronounced when $\sigma$ is large and $\mu$ is small. Similar to the bias for $\mu$, the bias for the 1/2 DL method when estimating $\sigma$ is variable with positive bias in low censoring scenarios and negative bias in higher censoring scenarios. Overall the average bias for the tcensReg method was generally closest to the gold standard with only slight overestimates of $\sigma$. However, we note that again the ratio of censoring to truncation appears to play an important role when comparing the Tobit and tcensReg methods. The Tobit method slightly outperforms the tcensReg method for scenarios with $\sigma=0.4$. In these settings the censoring to truncation ratio is higher as shown in Table~\ref{Tab:cens_pct_singlemean}, meaning that censoring is expected to occur much more frequently than truncation.

\begin{figure}[ht]
\centering
\caption{Average Log Mean Squared Error for $\sigma$ from Six Different Estimation Methods in Single Mean Model}
\includegraphics[width = \textwidth, keepaspectratio]{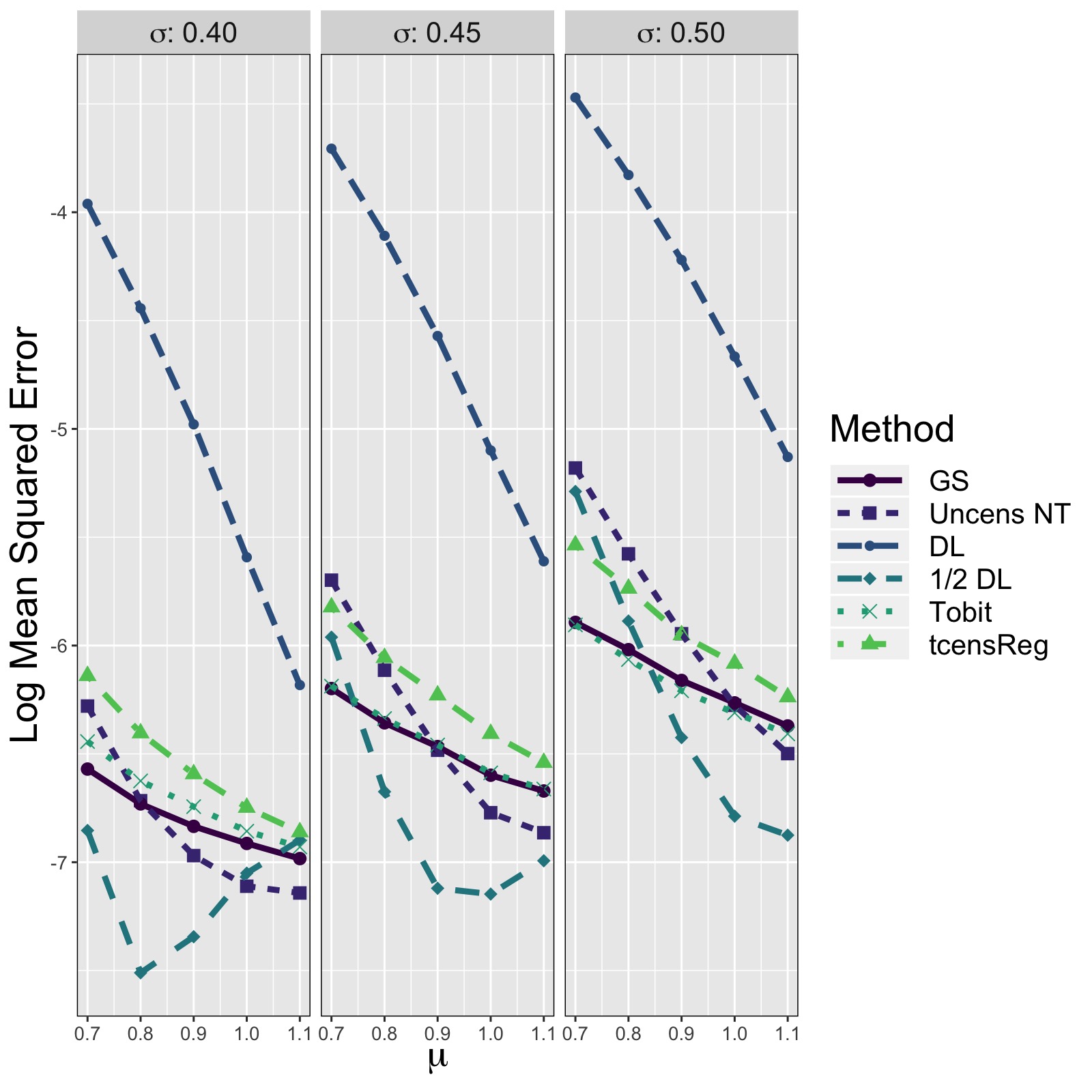}
\label{fig:tnorm_sd_mse}
\caption*{GS = Gold Standard, i.e. uncensored observations with truncation adjustment

Uncens NT = Uncensored data with no truncation adjustment

DL = detection limit

Tobit = Tobit censored regression with no truncation adjustment

tcensReg = Censored regression with truncation adjustment}
\end{figure}

Figure~\ref{fig:tnorm_sd_mse} shows the log MSE for each method. Again, the tcensReg method protected against false precision, with log MSE slightly greater than that of the gold standard. The 1/2 DL method had artificially low MSE for almost all parameter scenarios, while the Uncens NT alternated between over- and underestimation. The Tobit method was often the closest method to the gold standard.

\begin{figure}[ht]
\centering
\caption{Average Bias for $\delta$ from Six Different Estimation Methods in Two Population Model}
\includegraphics[width = \textwidth, keepaspectratio]{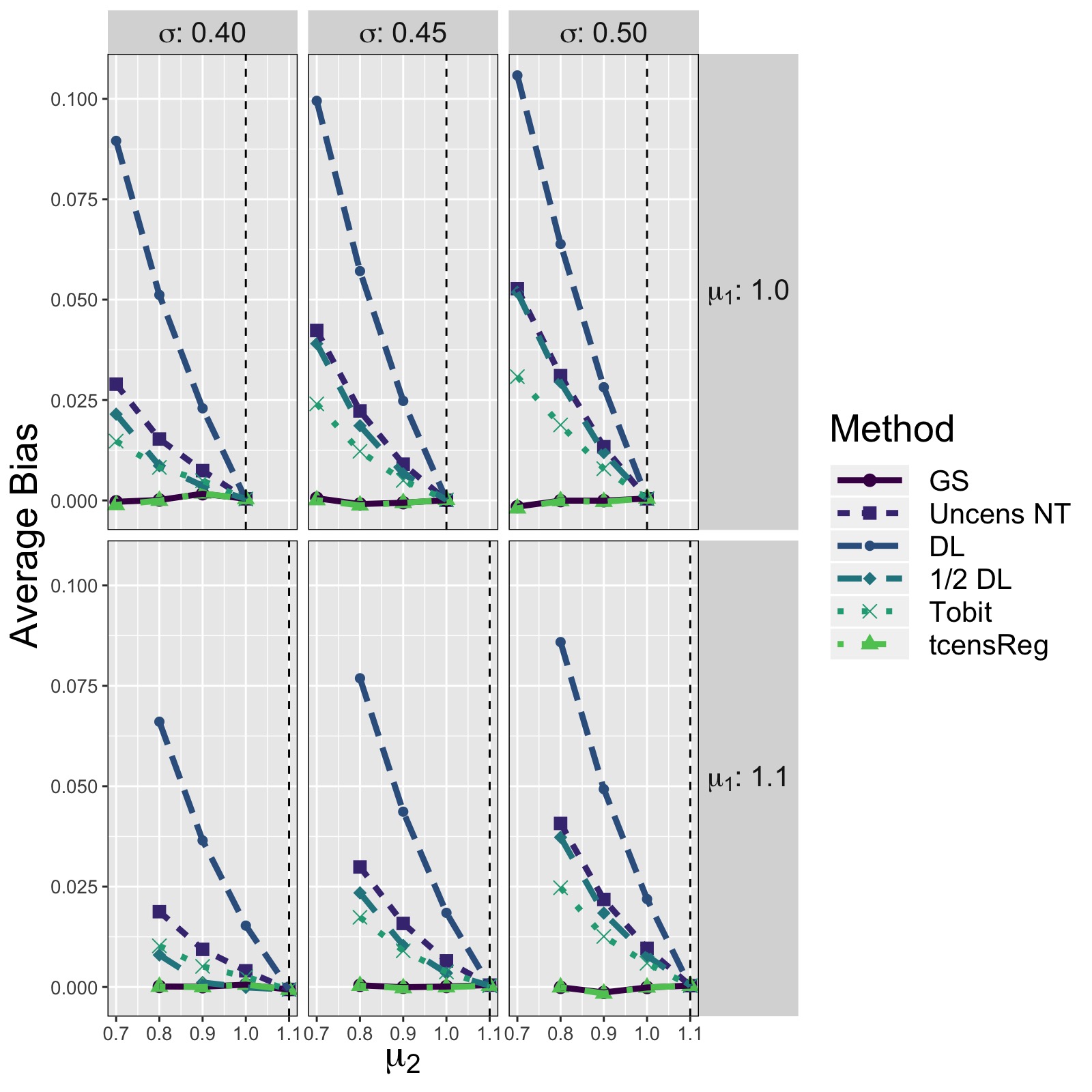}
\label{fig:tnorm_bias_diff}
\caption*{The vertical dashed black line corresponds to the case when $\delta=0$, i.e., $\mu_1=\mu_2$.

GS = Gold Standard, i.e. uncensored observations with truncation adjustment

Uncens NT = Uncensored data with no truncation adjustment

DL = detection limit

Tobit = Tobit censored regression with no truncation adjustment

tcensReg = Censored regression with truncation adjustment}
\end{figure}

In the two population simulation study, the parameter of primary interest was the difference of the means, $\delta$. Figure~\ref{fig:tnorm_bias_diff} shows the performance of each method with respect to average bias for $\delta$. For all values of $\delta$, the tcensReg and gold standard methods were essentially unbiased. The other four methods all had positive bias, corresponding to underestimation of the true difference in means. Similar to the single mean model, the greatest difference between the tcensReg method and the other methods occurred when the amount of censoring was greater, i.e., when $\mu_{1}=1.0$ and $|\delta|$ is large for a fixed $\sigma$.

For all six methods, the average bias of $\hat{\delta}$ when $\delta=0$ was approximately 0. As our simulation study for the single mean model showed, bias generally increases as the mean decreases for fixed $\sigma$; see Figure~\ref{fig:tnorm_mu_bias}. When the two means were equal, i.e., $\delta = 0$, the biases for estimating each mean were also equal and thus there was no bias for estimating $\delta$. 

\begin{figure}[ht]
\centering
\caption{Average Log Mean Squared Error for $\delta$ from Six Different Estimation Methods in Two Population Model}
\includegraphics[width = \textwidth, keepaspectratio]{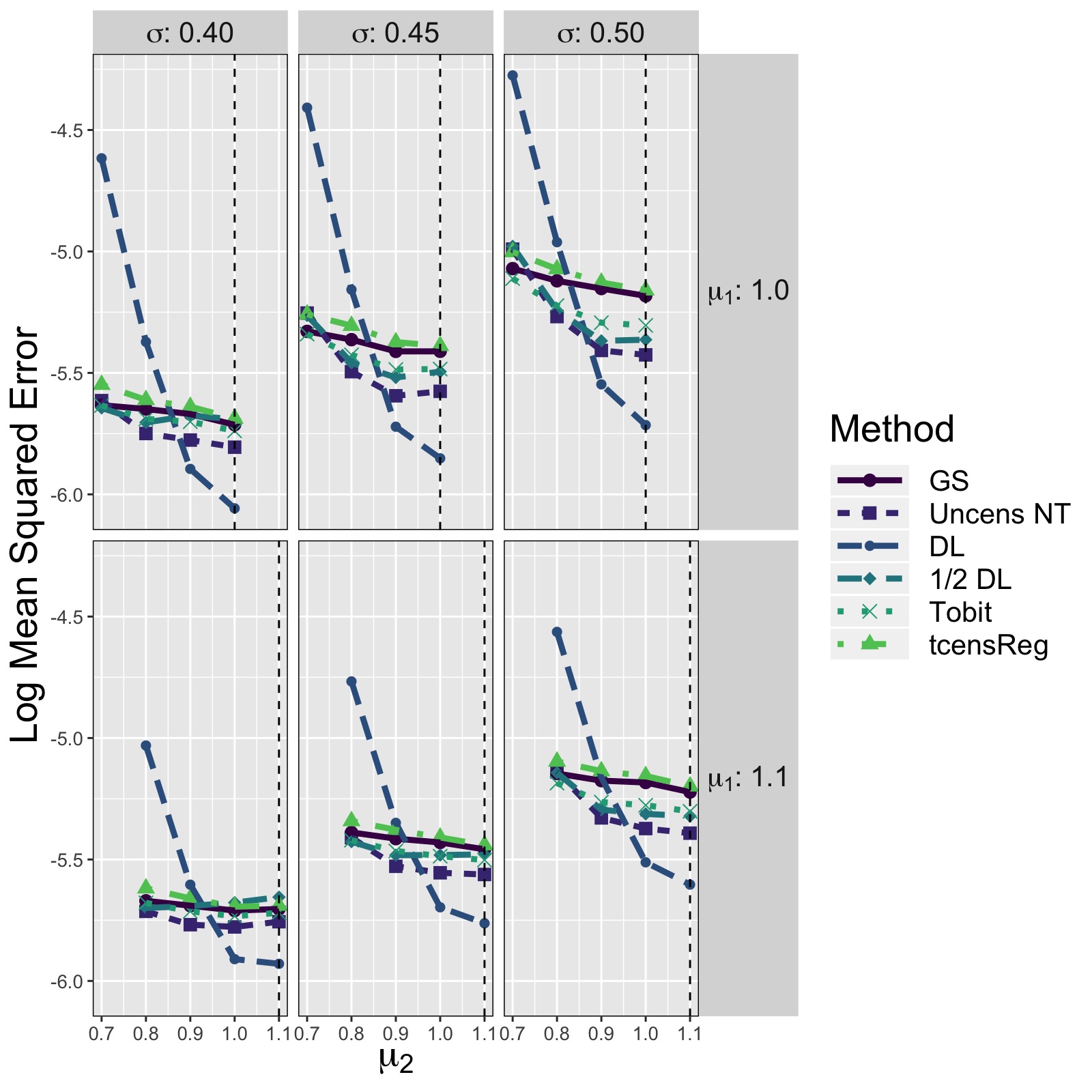}
\label{fig:tnorm_ssb_diff}\caption*{The vertical dashed black line corresponds to the case when $\delta=0$, i.e., $\mu_1=\mu_2$.

GS = Gold Standard, i.e. uncensored observations with truncation adjustment

Uncens NT = Uncensored data with no truncation adjustment

DL = detection limit

Tobit = Tobit censored regression with no truncation adjustment

tcensReg = Censored regression with truncation adjustment}
\end{figure}

The results for log MSE in the two-population simulation study (Figure~\ref{fig:tnorm_ssb_diff}) show that tcensReg had log MSE close to the gold standard for all scenarios. All of the other methods, with the exception of DL, had log MSE below that of the gold standard, indicating false precision. The DL method had high log MSE when $\delta$ was large (and its estimates were highly biased), and artificially low log MSE when $\delta$ was small. Overall, the tcensReg method was more successful than the other methods in capturing the true variability in the estimation procedure. 

\begin{figure}[ht]
\centering
\caption{Average Bias for Common $\sigma$ from Six Different Estimation Methods in Two Population Model}
\includegraphics[width = \textwidth, keepaspectratio]{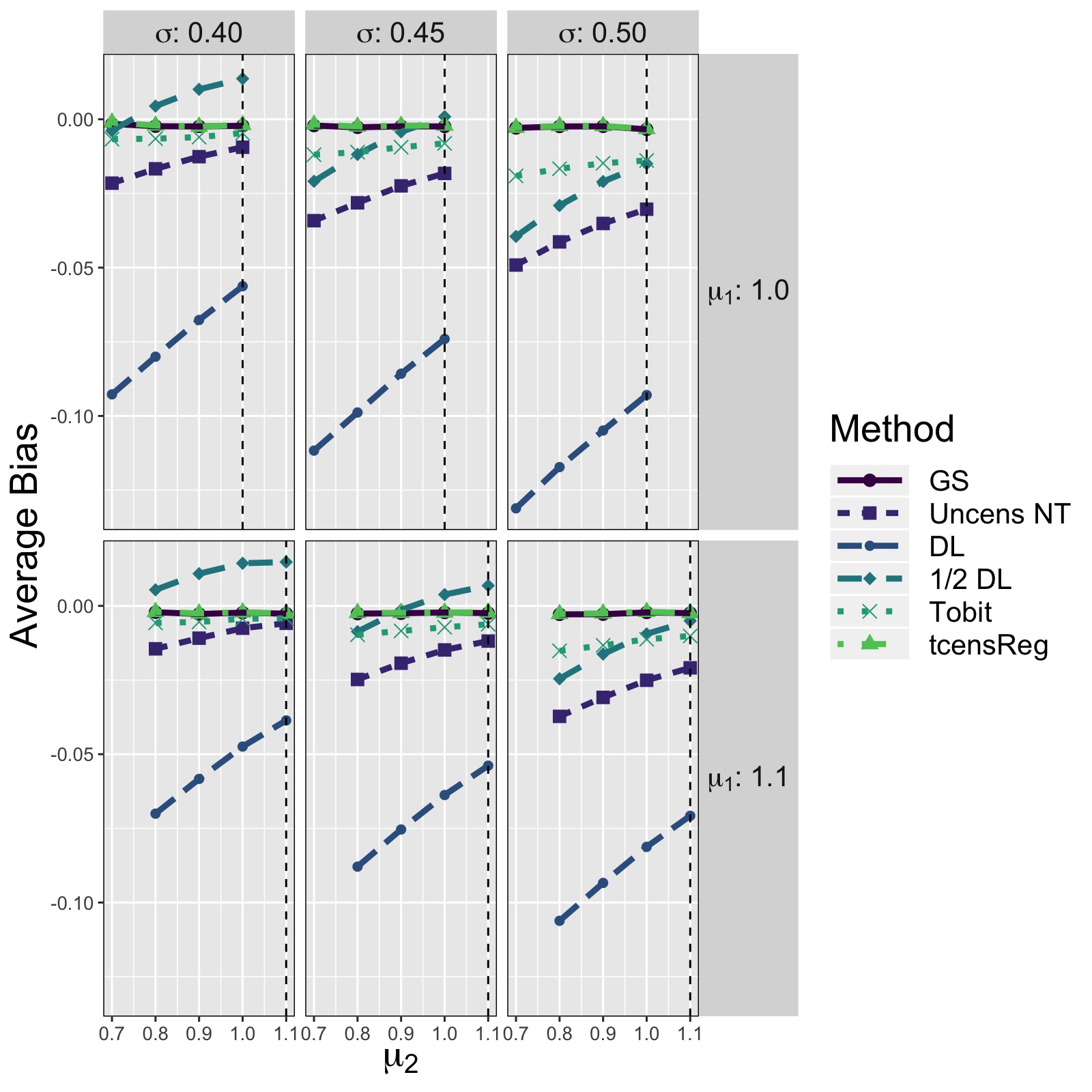}
\label{fig:tnorm_est_sd_common}
\caption*{The vertical dashed black line corresponds to the case when $\delta=0$, i.e., $\mu_1=\mu_2$.

GS = Gold Standard, i.e. uncensored observations with truncation adjustment

Uncens NT = Uncensored data with no truncation adjustment

DL = detection limit

Tobit = Tobit censored regression with no truncation adjustment

tcensReg = Censored regression with truncation adjustment}
\end{figure}

\begin{figure}[ht]
\centering
\caption{Average Log Mean Squared Error for Common $\sigma$ from Six Different Estimation Methods in Two Population Model}
\includegraphics[width = \textwidth, keepaspectratio]{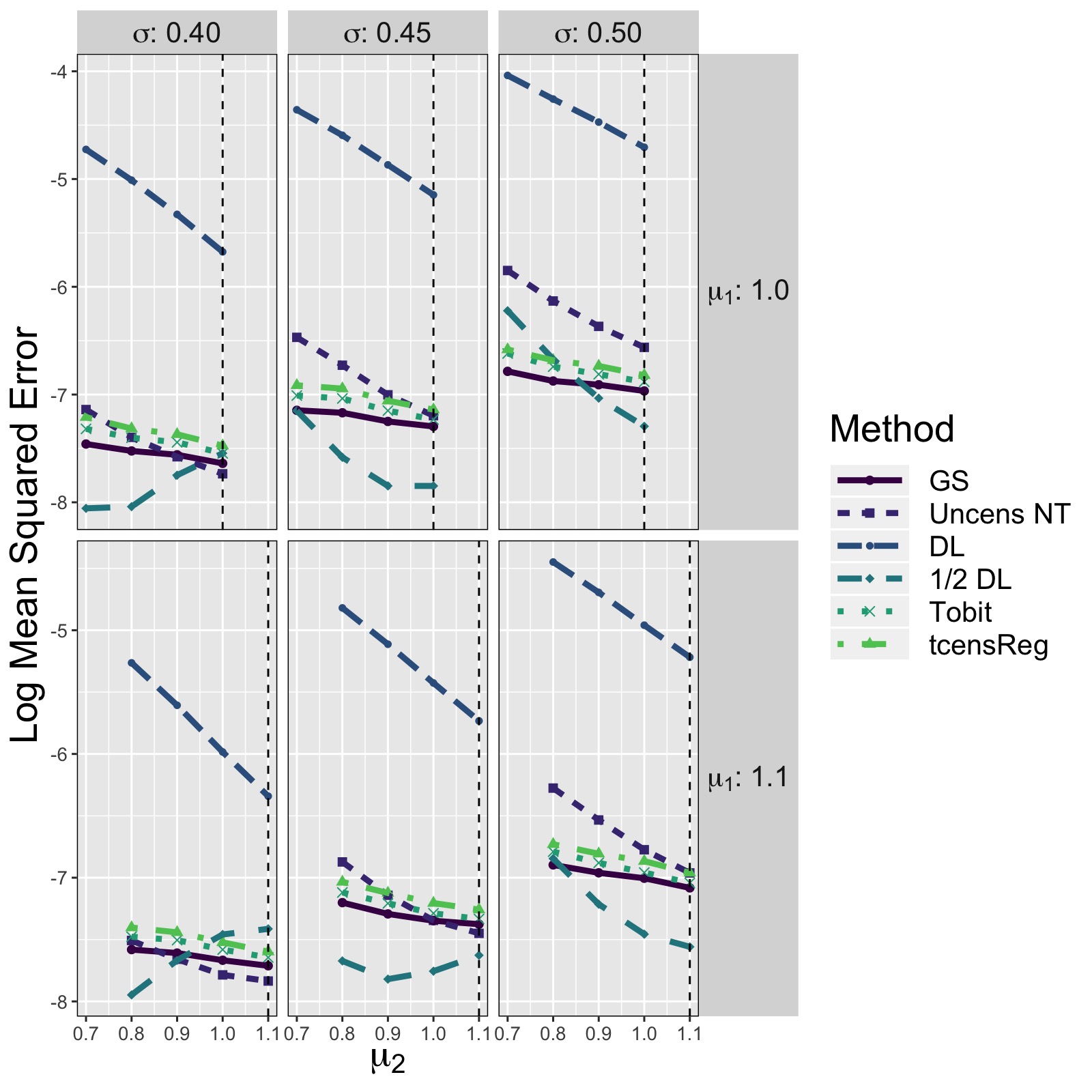}
\label{fig:tnorm_mse_sd_common}\caption*{The vertical dashed black line corresponds to the case when $\delta=0$, i.e., $\mu_1=\mu_2$.

GS = Gold Standard, i.e. uncensored observations with truncation adjustment

Uncens NT = Uncensored data with no truncation adjustment

DL = detection limit

Tobit = Tobit censored regression with no truncation adjustment

tcensReg = Censored regression with truncation adjustment}
\end{figure}

Also of interest in the two-population model is the estimate for the common standard deviation, $\sigma$. Figure~\ref{fig:tnorm_est_sd_common} shows the average bias of $\hat{\sigma}$ for each method. Similar to the single mean model, all methods other than tcensReg tended to underestimate the true standard deviation, especially when $\sigma$ was large. An exception was the 1/2 DL method, which alternated between over- and underestimating the variability of $\sigma$ but generally performed worse as $|\delta|$ increased. Figure~\ref{fig:tnorm_mse_sd_common} shows that both the tcensReg and Tobit methods had similar precision in the two population model with respect to the estimation of $\sigma$, while the other methods, i.e., DL, 1/2 DL and Uncens NT generally performed considerably worse. Overall in the two-population study, the tcensReg method had lower average bias in estimating $\delta$ and $\sigma$ along with consistent precision near the gold standard without the pitfalls of false precision.

Results for testing the non-inferiority hypothesis with $\alpha = 0.05$ are shown in Table~\ref{Tab:false_reject}. For all scenarios, the tcensReg method had slightly higher Type I error rates than the gold standard and the Tobit method had slightly higher Type I error rates than the tcensReg method. The DL method had Type I error rates 2.5-4 times higher than the nominal value. The other single imputation method, 1/2 DL, had inflated Type I errors in the medium and high variance settings. Especially when censoring was higher, i.e., $\mu_{1}=1.0$ and $\mu_{2}=0.85$, the other methods had significantly higher Type I error rates compared to tcensReg.

\section{Application}\label{application}
We now apply the methods to our contrast sensitivity application introduced in Section~\ref{example}. The goal was to compare visual quality, measured as contrast sensitivity, for monofocal vs multifocal lenses implanted following cataract surgery, using data collected from two clinical trials. Table~\ref{app_sum_table} shows the number of participants who received each type of IOL and the percent of patients whose observation was censored under each glare condition. Each group has over 150 patients; Multifocal 2 is larger than the others due to a 2:1 allocation ratio in that trial. As mentioned previously, the amount of censoring is greater with glare than without glare, with an average difference of 9\% censoring. However, all groups have relatively high levels of censoring, ranging from 11\%-29\%. 

\begin{figure}[ht]
\centering
\caption{95\% Confidence Intervals for Separate Standard Deviation for Monofocal vs Multifocal Lens at 12 CPD}
\includegraphics[width = 0.75\textwidth, keepaspectratio]{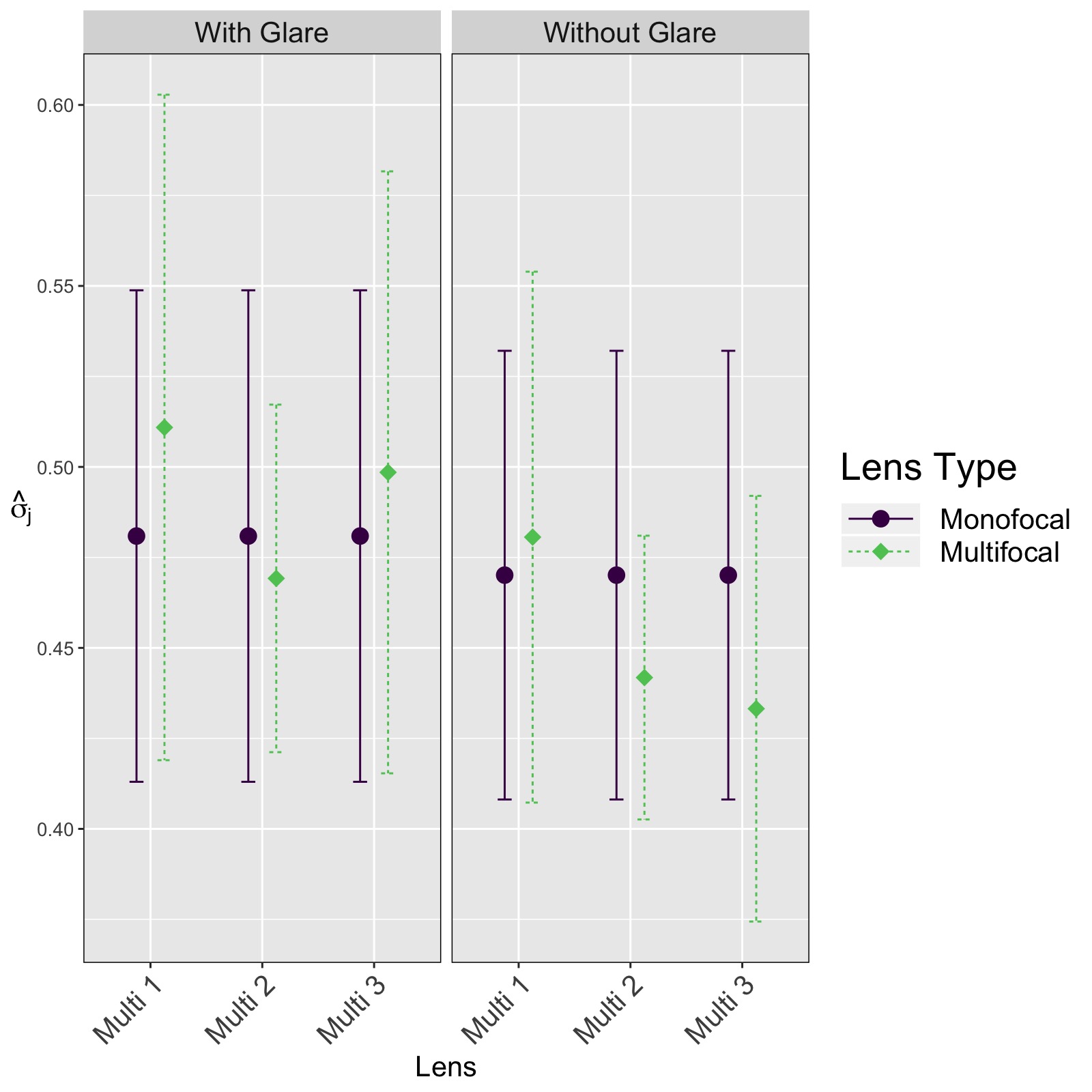}
\label{fig:sd_sepvar_app_tcensReg}
\end{figure}

Separate models were fit for each pairwise comparison between the monofocal lens and each multifocal lens with and without glare. Initially, each pairwise comparison was fit assuming the standard deviation for lens type was heteroskedastic. Figure~\ref{fig:sd_sepvar_app_tcensReg} displays result for the 95\% confidence interval of each groups standard deviation, $\hat{\sigma}_{j}$. The standard deviations for the multifocal lenses are sufficiently close to the monofocal lens for both glare conditions with a high level of overlap in the confidence intervals. This suggested that a common standard deviation for lens type is appropriate. In the subsequent analysis, each model was re-run assuming a common standard deviation in the two groups. The parameters of interest included the mean of each population (monofocal or multifocal), the mean difference between populations, and the common standard deviation. We compared four methods to estimate the parameters. DL imputation for censored observations, 1/2 DL imputation, Tobit regression and our censored truncated method, tcensReg.  

We also tested whether the multifocal lens is non-inferior to the monofocal lenses with respect to contrast sensitivity. Based on regulatory guidelines, the non-inferiority margin for contrast sensitivity is set at -0.15 log units \citep{iso_tech_report}. The non-inferiority test was one-sided test with $\alpha=0.05$, conducted by constructing a two-sided 90\% confidence interval (CI) for $\delta$ and comparing the lower bound of the CI to the non-inferiority margin. To establish non-inferiority, the lower bound for the CI must be above the non-inferiority margin. 

\begin{figure}[ht]
\centering
\caption{90\% Confidence Intervals for Difference in Monofocal vs Multifocal Lens at 12 CPD}
\includegraphics[width = 0.75 \textwidth, keepaspectratio]{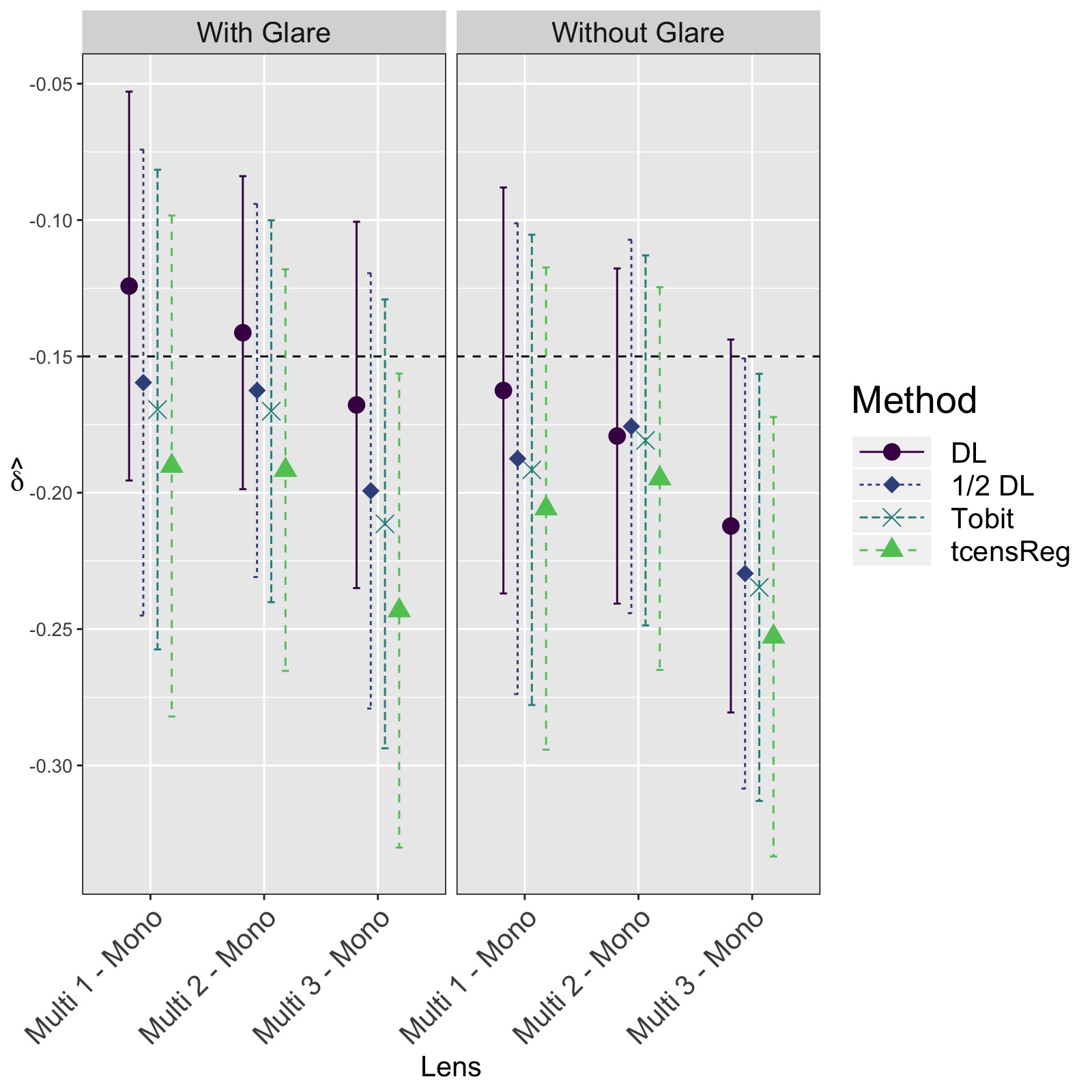}
\label{fig:ci_comparison}
\caption*{The horizontal dashed line at $\delta=-0.15$ indicates the non-inferiority margin.

DL = detection limit

Tobit = Tobit censored regression with no truncation adjustment

tcensReg = Censored regression with truncation adjustment}
\end{figure}

Figure~\ref{fig:ci_comparison} shows 90\% CIs for $\delta$ using the four methods. For all of the methods, the lower bound of the CIs are below the non-inferiority margin, for all three lens comparisons and both glare conditions. Thus using any of these methods, we would be unable to conclude that a particular multifocal lens is non-inferior to the monofocal lens in terms of contrast sensitivity. For each comparison, the tcensReg method had the largest estimate of $|\delta|$ and the DL method had the smallest estimate. The point estimates of $\delta$ from the Tobit and 1/2 DL methods were intermediate and similar.

\begin{figure}[ht]
\centering
\caption{Estimate of Common Standard Deviation in Monofocal vs Multifocal Lens at 12 CPD}
\includegraphics[width = 0.75\textwidth, keepaspectratio]{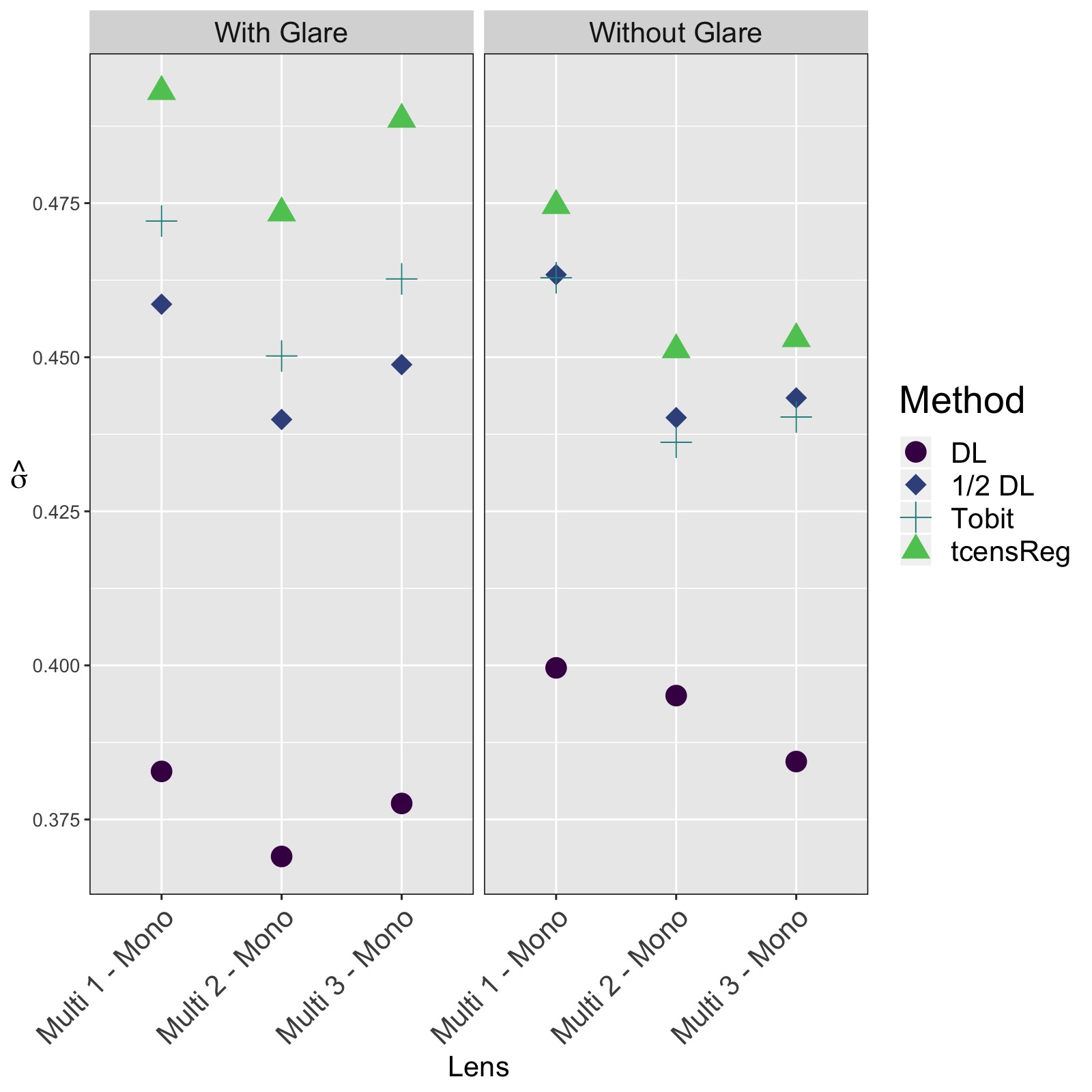}
\label{fig:sd_app_tcensReg}
\caption*{DL = detection limit

Tobit = Tobit censored regression with no truncation adjustment

tcensReg = Censored regression with truncation adjustment}
\end{figure}

The tcensReg method yielded the longest CIs (mean length~=~0.164 log units). Mean CI lengths for the Tobit, 1/2 DL and DL methods were 0.158, 0.156 and 0.133 log units, respectively. The differences in CI length are a reflection of differences in estimates of $\sigma$, shown in Figure~\ref{fig:sd_app_tcensReg}. The DL method had consistently low estimates of $\sigma$, while the tcensReg method had the highest estimated values of $\sigma$. The 1/2 DL and Tobit methods gave similar estimates intermediate between the other two methods. 

Interpreting these results in light of our simulation studies, if the latent normality assumption holds for these data, we would expect the estimates from the tcensReg method to have the least bias. The other methods typically underestimated both the difference in means and the population standard deviation, which would tend to lead to confidence intervals that are more likely to falsely exclude the non-inferiority margin and thus to higher Type I error rates, especially when there is a high rate of censoring. The wider confidence intervals observed in Figure~\ref{fig:ci_comparison} from the tcensReg may better protect against Type I errors.

\section{Discussion}\label{discussion}
In this manuscript, we developed a maximum likelihood method for estimating parameters from a truncated normal distribution when observations are subject to censoring. We have also developed the R package tcensReg to implement the method. We showed in simulations that our method has substantially less bias for estimating the mean and standard deviation of the underlying latent normal distribution than other commonly used methods for a range of simulation settings. Our method also had close to the nominal Type I error rate for non-inferiority testing, while single imputation with either the detection limit or half the detection limit and Tobit regression often had inflated Type I error rates when the censoring rate was high.

The single mean simulation study showed that as the levels of censoring and truncation increased, the bias for all other methods generally increased, often dramatically. As expected, point imputation of the detection limit was consistently the worst performing method, but even the Tobit method had large bias under certain conditions. An important factor when comparing the Tobit and tcensReg method in the single mean model was not only the raw levels of censoring and truncation but the ratio of censoring to truncation. For low censoring to truncation ratios, generally below 20, the tcensReg method outperformed Tobit, while high ratios tended to have similar performance in mean estimation and worse performance in standard deviation estimation. This trend may point to the scenarios where the censoring dominates truncation in terms of estimation and thus using the Tobit method can lead to more precise estimates, particularly of the standard deviation. It is important to note that the tcensReg method was the only method to have positive bias in estimation of the standard deviation while other methods were typically negatively biased. Positive bias in this sense leads to conservative hypothesis testing.

In the two population simulation study, we observed trends similar to those for the single mean model. The tcensReg method uniformly outperformed all other methods in estimating the average difference and standard deviation between the populations with respect to average bias. The greatest difference between the methods occurred when the censoring rate was high and the difference between the population means was large. Unlike the single mean model, the ratio of censoring to truncation did not appear to play a significant role in the accuracy of parameter estimation for the two-population method. Similar to the single mean model, the precision of alternative methods was also inaccurate, leading to values of the mean squared error that appeared to outperform the gold standard. Across all parameter scenarios, the tcensReg method had significantly lower average bias and did not fall victim to the false precision fallacy. 

Results from the non-inferiority simulation study confirmed the trends observed in the previous simulation studies. Methods other than tcensReg tended to underestimate the true variability of the underlying data generating mechanism, which led to precise but biased results. This bias tended to increase as the censoring increased, and this combined with underestimated values of the true standard deviation led to substantially inflated Type I error rates. The tcensReg method consistently had estimates closest to the gold standard.

In the application, the main difference among the methods was a greater estimate of the difference between means when using the tcensReg method. The tcensReg method also had the widest confidence interval length due to higher estimates of the variance. The DL method results suggested that it would be the most likely method to lead to a finding of non-inferiority, but as shown in the simulation study, this method can have highly inflated Type I error rates. The 1/2 DL, Tobit, and tcensReg methods are more likely to capture the true difference and variability in the estimate, with the tcensReg method showing evidence that the true difference may be even greater than previously thought using either of the other two methods. In our particular application, none of these methods differed with respect to the ultimate conclusion of the non-inferiority hypothesis test. However, in cases in which non-inferiority is more marginal, the choice of method could make a difference. 

The goal of the estimation methods presented here is to estimate the mean and standard deviation of the latent normal distribution rather than the mean and standard deviation of the truncated normal distribution. This affects the interpretation of the results. When censoring is less than 50 percent, the mean of the latent normal corresponds to the mode of the truncated normal. In many settings the mode of the truncated normal may be a more clinically meaningful measure of central tendency than the mean of the truncated normal. For example, in contrast sensitivity testing, it may be clinically relevant to understand what factors affect the mode rather than the mean in order to target where the greatest proportion of patients lie.

The methodology as presented makes a strong parametric assumption of normality. For studies with small to moderate sample sizes, checking the reasonableness of the normality assumption may be difficult and the assumption may be increasingly tenuous as the amount of censoring and truncation increase. While the unobserved data may not be exactly normal, a normality assumption at least reflects an assumption that the unobserved part of the distribution has a monotone decreasing shape, which is frequently a reasonable assumption. In general, whenever an analysis involves unverifiable assumptions, conducting a sensitivity analysis is prudent. The tcensReg method can be viewed as an option in the toolkit of the statistician that can be used as part of a sensitivity analysis, which might also include Tobit regression.

Another potential limitation is that the detection threshold and truncation value are assumed to be known. While assuming that the detection threshold is known is often reasonable, the truncation value is not observable in the data due to the censoring and so must be incorporated in the analysis based on subject matter expertise. 

In this manuscript, we have focused on the setting of left censoring and left truncation. The methods can be easily extended to handle right censoring with right truncation. Future work could extend the methods to handle censoring in both tails of the distribution. In our application, we noted some suggestion of censoring due to an upper detection threshold; see Figure~\ref{fig:marg_hist}. Conducting the analysis adjusting for left and right censoring could potentially improve the accuracy of the estimates. 

Other possible extensions include extension to other non-normal parametric distributions. Also, this framework for estimation could be extended to linear mixed effect models with repeated measurements using the extended Newton-Raphson technique proposed by \cite{lindstrom1988newton}.

\section*{Acknowledgments}
This methodology was developed in part during an internship program at Alcon Laboratories, Inc. (Fort Worth, TX, USA). Thank you to the Biostatistics team at Alcon for their feedback and suggestions during the development of this manuscript. R code for generating data used in the simulations discussed in Section~\ref{simulation} is available at \url{https://github.com/williazo/tcensReg/tree/master/inst/script}. Data used for the application in Section~\ref{application} is proprietary to Alcon Laboratories, Inc.%

\subsection*{Financial disclosure}

None reported.

\subsection*{Conflict of interest}
Hyung-Woo Kim is a former employee of Alcon Laboratories, Inc.

\clearpage
\bibliographystyle{DeGruyter}
\bibliography{tcensReg_cit}

\clearpage
\section*{Tables and Figures}\label{tab_figs}
\begin{table}[!ht]
\caption{Manufacturer Log Contrast Sensitivity Value$^1$}
\label{Tab:vv_logval}
\centering
\begin{threeparttable}[h]
\begin{tabular}{| c | c | c | c | c | c | c | c | c | c | c |}
	\hline 
	& & \multicolumn{9}{c|}{\textbf{Contrast Sensitivity Threshold Scores}}\\
	\hline
	& & 0 & 1 & 2 & 3 & 4 & 5 & 6 & 7 & 8\\
	\hline
\multirow{5}{*}{\textbf{CPD$^2$}} & \textbf{1.5} & 0.60 & 0.90 & 1.07 & 1.22 & 1.37 & 1.52 & 1.67 & 1.82 & 1.97\\
	& \textbf{3.0} & 0.70 & 1.00 & 1.17 & 1.34 & 1.49 & 1.63 & 1.78 & 1.93 & 2.08 \\
	& \textbf{6.0} & 0.91 & 1.21 & 1.38 & 1.55 & 1.70 & 1.84 & 1.99 & 2.14 & 2.29 \\
	& \cellcolor{gray}  \textbf{12.0} & \cellcolor{gray} \fcolorbox{red}{gray}{0.61} & \cellcolor{gray} 0.91 & \cellcolor{gray} 1.08 & \cellcolor{gray} 1.25 & \cellcolor{gray} 1.40 & \cellcolor{gray} 1.54 & \cellcolor{gray} 1.69 & \cellcolor{gray} 1.84 & \cellcolor{gray} 1.99 \\
	& \textbf{18.0} & 0.17 & 0.47 & 0.64 & 0.81 & 0.96 & 1.10 & 1.25 & 1.40 & 1.55 \\
	\hline
\end{tabular}
\begin{tablenotes}
\footnotesize
	\item[1] Based on scoring instructions from \url{http://www.vectorvision.com/csv1000-norms/} accessed on 08NOV2018
	\item[2] CPD = Cycles per Degree
	\item Data analyzed in Section~\ref{application} is from 12.0 CPD test highlighted in grey
	\item Detection limit (DL) for 12.0 CPD is 0.61 highlighted in the red box and therefore 1/2 DL is 0.31
\end{tablenotes}
\end{threeparttable}
\end{table}

\clearpage
\begin{table}[!ht]
\caption{Expected Percentages of Truncation and Censoring in Single Mean Simulation Study}
\label{Tab:cens_pct_singlemean}
\centering
\begin{threeparttable}
\begin{tabular}{|c|c|c|c|c|} 
	\hline
	$\mu$ & $\sigma$ &  Expected Censoring \% &  Expected Truncated \% & Censoring:Truncation Ratio$^*$\\
	\hline
	\multirow{3}{*}{1.1} & 0.50 & 15.17\% & 1.39\% & 10.91\\
	& 0.45 & 13.18\% & 0.73\% & 18.05\\
	& 0.40 & 10.76\% & 0.30\% & 35.87\\
	\hline
	\multirow{3}{*}{1.0} & 0.50 & 19.95\% & 2.28\% & 8.75\\
	& 0.45 & 18.23\% & 1.31\% & 13.92\\
	& 0.40 & 15.96\% & 0.62\% & 25.74\\
	\hline
	\multirow{3}{*}{0.9} & 0.50 & 25.42\% & 3.59\% & 7.08\\
	& 0.45 & 24.24\% & 2.28\% & 10.63\\
	& 0.40 & 22.47\% & 1.22\% & 18.42\\
	\hline
	\multirow{3}{*}{0.8} & 0.50 & 31.44\% & 5.48\% & 5.74\\
	& 0.45 & 31.04\% & 3.77\% & 8.23\\
	& 0.40 & 30.15\% & 2.28\% & 13.22\\
	\hline
	\multirow{3}{*}{0.7} & 0.50 & 37.84\% & 8.08\% & 4.68\\
	& 0.45 & 38.38\% & 5.99\% & 6.41\\
	& 0.40 & 38.64\% & 4.01\% & 9.64\\
	\hline
\end{tabular}
\begin{tablenotes}
	\item Note that censoring threshold set at $\nu = 0.61$ and truncation value at $a=0$.
	\item[*] This column compares the expected percent of censoring to the expected percent of truncation, e.g., when $\mu$ = 1.1 and $\sigma$ = 0.50 we expect 10.91 times more censoring than truncation.
\end{tablenotes}
\end{threeparttable}
\end{table}

\clearpage
\begin{table}[!ht]
\caption{Type I Error Rates for Non-inferiority Test in Simulation Study}
\label{Tab:false_reject}
\centering
\begin{threeparttable}
\begin{tabular}{cccccccccc}
  \hline
 $\mu_{1}$ & $\mu_2$ & $\delta$ & $\sigma$ & GS$^1$ & Uncens NT$^2$ & DL$^3$ & 1/2 DL & Tobit$^4$ & tcensReg$^5$ \\ 
 \hline
1.1 & 0.95 & -0.15 & 0.40 & 0.0552 & 0.0609 & 0.1240 & 0.0477 & 0.0576 & 0.0555 \\ 
1.0 & 0.85 & -0.15 & 0.40 & 0.0563 & 0.0709 & 0.1874 & 0.0577 & 0.0652 & 0.0594 \\ 
\hline
1.1 & 0.95 & -0.15 & 0.45 & 0.0606 & 0.0736 & 0.1424 & 0.0648 & 0.0712 & 0.0642 \\ 
1.0 & 0.85 & -0.15 & 0.45 & 0.0642 & 0.0863 & 0.2046 & 0.0766 & 0.0776 & 0.0682 \\ 
\hline
1.1 & 0.95 & -0.15 & 0.50 & 0.0582 & 0.0755 & 0.1456 & 0.0687 & 0.0670 & 0.0599 \\ 
1.0 & 0.85 & -0.15 & 0.50 & 0.0603 & 0.0887 & 0.1996 & 0.0831 & 0.0754 & 0.0641 \\ 
   \hline
\end{tabular}
\begin{tablenotes}
	\item \textbf{Note}: nominal $\alpha$ set to 0.05
	\item[1] GS = Gold Standard, i.e. uncensored observations with truncation adjustment
	\item[2] Uncens NT = Uncensored data with no truncation adjustment
	\item[3] DL = detection limit
	\item[4] Tobit = Tobit censored regression with no truncation adjustment
	\item[5] tcensReg = Censored regression with truncation adjustment
\end{tablenotes}
\end{threeparttable}
\end{table}

\clearpage
\begin{table}[!ht]
\caption{Observed Censoring Percentage for Intraocular Lenses (IOLs) under Dim Lighting at 12.0 CPD}
\label{app_sum_table}
\centering
\begin{tabular}{c  c  c  c }
	\hline 
	& & \multicolumn{2}{c}{\textit{Observed Censoring \%}}\\
	\textbf{IOL Type} & \textbf{n} & \textbf{With Glare} & \textbf{Without Glare}  \\
	\hline
	Monofocal$^*$ & 159 & 17.6\% & 12.6\% \\
	Multifocal 1$^*$ & 153 & 29.4\% & 20.9\% \\
	Multifocal 2$^{**}$ & 377 & 24.7\% & 11.4\% \\
	Multifocal 3$^{**}$ & 185 & 28.1\% & 18.4\% \\
	\hline
\end{tabular}
\caption*{$^*$ Data from \href{https://clinicaltrials.gov}{ClinicalTrials.gov} Identifier NCT01510717\\
	$^{**}$ Data from \href{https://clinicaltrials.gov}{ClinicalTrials.gov} Identifier NCT01424189}
\end{table}

\clearpage

\clearpage

\clearpage

\clearpage

\clearpage

\clearpage

\clearpage

\clearpage

\clearpage

\clearpage

\clearpage

\clearpage

\clearpage

\clearpage

\clearpage
\appendix
\section{Formula for the Gradient}\label{grad_simple}
In the model discussed in Section~\ref{lin_tcensReg}, the parameters to be estimated are $\boldsymbol{\theta} = (\boldsymbol{\beta}, \sigma)^{T}$ where $\boldsymbol{\beta}=(\beta_{1},\dots,\beta_{p-1})^{T}$ from Equation~\ref{eqn:log_like_lin}. The single mean model is a special case of the linear equation where $\mathbf{X}_{i}^{T}\boldsymbol{\beta}=\mu$. 

Investigating the log likelihood from Equation~\ref{eqn:log_like_lin}, it is evident that calculating the partial derivatives with respect to $\ln(\sigma)$ rather than $\sigma$ is optimal. Then to obtain estimates and standard errors for $\sigma$, one can apply the inverse transformation and delta method.

Note that another possible parametrization using one-to-one functions is $\delta=\frac{\beta}{\sigma}$ and $\eta=\frac{1}{\sigma}$ used by \cite{tobin1958estimation}. However, we present results for the derivation using $\beta$ and $\ln(\sigma)$ analogous to methods used by \cite{henningsen2010estimating} in the R package censReg.

Throughout this derivation, let $c_{i}^{*}=\frac{c-\mathbf{x}_{i}^{T}\boldsymbol{\beta}}{\sigma}$ for some constant $c$. The following properties will be used to calculate the appropriate derivatives:
\begin{itemize}
\item $\frac{\partial}{\partial x}\Phi(x)=\phi(x)$
\item $\frac{\partial}{\partial x}\phi(x)= -x\phi(x)$
\item $\frac{\partial}{\partial \ln(\sigma)}\sigma=\frac{\partial}{\partial \ln(\sigma)}\exp\big[\ln(\sigma)\big]=\sigma$
\item $\frac{\partial}{\partial \ln(\sigma)}c_{i}^{*}=-c_{i}^{*}$
\end{itemize}

We define the gradient as 
\begin{equation*}
\nabla l(\boldsymbol{\theta})=\begin{pmatrix}
\frac{\partial l}{\partial \beta_{1}}\\ \vdots \\ \frac{\partial l}{\partial \beta_{p-1}} \\[6pt]  \frac{\partial l}{\partial \ln(\sigma)}
\end{pmatrix}, \ \text{for} \ p\ge2.
\end{equation*}

Taking the derivative of the log-likelihood with respect to $\beta_{k}$,
\begin{equation}\label{eqn:d_betak}
\frac{\partial l}{\partial \beta_{k}}=-\sum_{i=1}^{n}\frac{x_{ik}\phi(a_{i}^{*})}{\sigma[1-\Phi(a_{i}^{*})]}
-\sum_{i\in S_{0}}\frac{x_{ik}\big[\phi(\nu_{i}^{*})-\phi(a_{i}^{*})\big]}{\sigma[\Phi(\nu_{i}^{*})-\Phi(a_{i}^{*})]}
+\frac{1}{\sigma^{2}}\sum_{i\in S_{1}}x_{ik}(y_{i}-\mathbf{x}_{i}^{T}\boldsymbol{\beta}) \ \ \text{for} \ k = 1,\dots, p-1.
\end{equation}

Then taking the derivative with respect to $\ln(\sigma)$,
\begin{equation}\label{eqn:d_logsigma}
\frac{\partial l}{\partial \ln(\sigma)}=-\sum_{i=1}^{n}\frac{a_{i}^{*}\phi(a_{i}^{*})}{1-\Phi(a_{i}^{*})}
-\sum_{i\in S_{0}}\frac{\big[\nu_{i}^{*}\phi(\nu_{i}^{*})-a_{i}^{*}\phi(a_{i}^{*})\big]}{\Phi(\nu_{i}^{*})-\Phi(a_{i}^{*})} -n_{1}
+\frac{1}{\sigma^{2}}\sum_{i\in S_{1}}(y_{i}-\mathbf{x}_{i}^{T}\boldsymbol{\beta})^{2}
\end{equation}

Therefore, the gradient vector is
\begin{equation}\label{eqn:gradient}
\nabla l(\boldsymbol{\theta})=\begin{pmatrix}
-\sum_{i=1}^{n}\frac{x_{i1}\phi(a_{i}^{*})}{\sigma[1-\Phi(a_{i}^{*})]}
-\sum_{i\in S_{0}}\frac{x_{i1}\big[\phi(\nu_{i}^{*})-\phi(a_{i}^{*})\big]}{\sigma[\Phi(\nu_{i}^{*})-\Phi(a_{i}^{*})]}
+\frac{1}{\sigma^{2}}\sum_{i\in S_{1}}x_{i1}(y_{i}-\mathbf{x}_{i}^{T}\boldsymbol{\beta}) \\ \vdots \\
-\sum_{i=1}^{n}\frac{x_{i(p-1)}\phi(a_{i}^{*})}{\sigma[1-\Phi(a_{i}^{*})]}
-\sum_{i\in S_{0}}\frac{x_{i(p-1)}\big[\phi(\nu_{i}^{*})-\phi(a_{i}^{*})\big]}{\sigma[\Phi(\nu_{i}^{*})-\Phi(a_{i}^{*})]}
+\frac{1}{\sigma^{2}}\sum_{i\in S_{1}}x_{i(p-1)}(y_{i}-\mathbf{x}_{i}^{T}\boldsymbol{\beta})\\[6pt]
-\sum_{i=1}^{n}\frac{a_{i}^{*}\phi(a_{i}^{*})}{1-\Phi(a_{i}^{*})}
-\sum_{i\in S_{0}}\frac{\big[\nu_{i}^{*}\phi(\nu_{i}^{*})-a_{i}^{*}\phi(a_{i}^{*})\big]}{\Phi(\nu_{i}^{*})-\Phi(a_{i}^{*})} -n_{1}
+\frac{1}{\sigma^{2}}\sum_{i\in S_{1}}(y_{i}-\mathbf{x}_{i}^{T}\boldsymbol{\beta})^{2}
\end{pmatrix}.
\end{equation}

\section{Formula for the Hessian}\label{hess_simple}
The Hessian is the matrix of second derivatives,
\begin{equation*}
\nabla^{2}l(\boldsymbol{\theta})=\begin{bmatrix}
\frac{\partial^{2} l}{\partial \beta_{1}^{2}} & \cdots & \frac{\partial^{2} l}{\partial \beta_{1} \partial\beta_{p-1}} & \frac{\partial^{2} l}{\partial \beta_{1} \partial\ln(\sigma)}\\[6pt]
  \vdots & \ddots  & \vdots & \vdots \\[6pt]
  \frac{\partial^{2} l}{\partial\beta_{p-1} \partial \beta_{1}}& \cdots & \frac{\partial^{2} l}{\partial \beta_{p-1}^{2}} &  \frac{\partial^{2} l}{\partial \beta_{p-1} \partial \ln(\sigma)}\\[6pt]
  \frac{\partial^{2} l}{\partial\ln(\sigma) \partial \beta_{1}} & \cdots & \frac{\partial^{2} l}{\partial \ln(\sigma) \partial \beta_{p-1}} &  \frac{\partial^{2} l}{\partial \ln^{2}(\sigma)}
\end{bmatrix}.
\end{equation*}
Note that this Hessian matrix is symmetric so that $\nabla^{2}l(\boldsymbol{\theta})_{ij}=\nabla^{2}l(\boldsymbol{\theta})_{ji}$ for $i\ne j$.

The individual components of this matrix are calculated as
\begin{equation}\label{eqn:d_betak_betal}
\scriptsize
\begin{split}
\frac{\partial^{2} l}{\partial \beta_{k}\partial\beta_{l}}=&
-\sum_{i=1}^{n}\frac{x_{ik}x_{il}\Big\{a_{i}^{*}[1-\Phi(a_{i}^{*})]\phi(a_{i}^{*})-\phi^{2}(a_{i}^{*})\Big\}}{\sigma^{2}[1-\Phi(a_{i}^{*})]^{2}}\\
&-\sum_{i\in S_{0}}\frac{x_{ik}x_{il}\Big\{[\nu_{i}^{*}\phi(\nu_{i}^{*})-a_{i}^{*}\phi(a_{i}^{*})][\Phi(\nu_{i}^{*})-\Phi(a_{i}^{*})]+[\phi(\nu_{i}^{*})-\phi(a_{i}^{*})]^{2}\Big\}}{\sigma^{2}[\Phi(\nu_{i}^{*})-\Phi(a_{i}^{*})]^{2}}\\
&-\frac{1}{\sigma^{2}}\sum_{i\in S_{1}}x_{ik}x_{il}\ \text{for} \ k= 1,\dots,p-1 \ \text{and} \ l = k,\dots,p-1,
\end{split}
\end{equation}

\begin{equation}\label{eqn:d_betak_logsigma}
\scriptsize
\begin{split}
\frac{\partial^{2} l}{\partial \beta_{k} \partial \ln(\sigma)}=&
\sum_{i=1}^{n}\frac{x_{ik}\Big\{[1-\Phi(a_{i}^{*})]\phi(a_{i}^{*})[1-(a_{i}^{*})^{2}]+a_{i}^{*}\phi^{2}(a_{i}^{*})\Big\}}{\sigma[1-\Phi(a_{i}^{*})]^{2}}\\
&-\sum_{i\in S_{0}} \frac{x_{ik}\bigg\{\Big(\Phi(\nu_{i}^{*})-\Phi(a_{i}^{*})\Big)\Big(\phi(\nu_{i}^{*})\big[1-(\nu_{i}^{*})^{2}\big]-\phi(a_{i}^{*})\big[1-(a_{i}^{*})^{2}\big]\Big)-\Big[\phi(\nu_{i}^{*})-\phi(a_{i}^{*})\Big]\Big[\phi(\nu_{i}^{*})\nu_{i}^{*}-\phi(a_{i}^{*})a_{i}^{*}\Big]\bigg\}}{\sigma[\Phi(\nu_{i}^{*})-\Phi(a_{i}^{*})]^{2}}\\
&-\frac{2}{\sigma^{2}}\sum_{i\in S_{1}}x_{ik}(y_{i}-\mathbf{x}_{i}^{T}\boldsymbol{\beta}) \ \text{for} \ k = 1,\dots,p-1,
\end{split}
\end{equation}

\begin{equation}\label{eqn:d_logsigma2}
\scriptsize
\begin{split}
\frac{\partial^{2} l}{\partial \ln^{2}(\sigma)}=&
\sum_{i=1}^{n}\frac{a_{i}^{*}\Big\{[1-\Phi(a_{i}^{*})]\phi(a_{i}^{*})[1-(a_{i}^{*})^{2}]+a_{i}^{*}\phi^{2}(a_{i}^{*})\Big\}}{[1-\Phi(a_{i}^{*})]^{2}}\\
&-\sum_{i\in S_{0}}\frac{\Big(\Phi(\nu_{i}^{*})-\Phi(a_{i}^{*})\Big)\Big(\phi(\nu_{i}^{*})\big[(\nu_{i}^{*})^{3}-\nu_{i}^{*}\big]-\phi(a_{i}^{*})\big[(a_{i}^{*})^{3}-a_{i}^{*}\big]\Big)+\Big[\phi(\nu_{i}^{*})\nu_{i}^{*}-\phi(a_{i}^{*})a_{i}^{*}\Big]^{2}}
{[\Phi(\nu_{i}^{*})-\Phi(a_{i}^{*})]^{2}}\\
&-\frac{2}{\sigma^{2}}\sum_{i\in S_{1}}(y_{i}-\mathbf{x}_{i}^{T}\boldsymbol{\beta})^{2}.
\end{split}
\end{equation}

\section{Gradient for Heteroskedastic Model}\label{grad_sepvar}
Assuming that there are samples from $J$ independent populations as discussed in Section~\ref{lin_tcensReg}, the parameters to be estimated are $\boldsymbol{\theta} = (\boldsymbol{\beta}, \sigma_{1},\dots,\sigma_{J})^{T}$ where $\boldsymbol{\beta}=(\beta_{1},\dots,\beta_{p-1})^{T}$. Again, the form of log likelihood from Equation~\ref{eqn:log_like_sepvar} suggests calculating the partial derivatives with respect to $\ln(\sigma_{j})$ rather than $\sigma_{j}$ is optimal. Let $c_{ij}^{*}=\frac{c-\mathbf{x}_{ij}^{T}\boldsymbol{\beta}}{\sigma_{j}}$ for some constant $c$.

We define the gradient as 
\begin{equation*}
\nabla l(\boldsymbol{\theta})=\begin{pmatrix}
\frac{\partial l}{\partial \beta_{1}}\\
\vdots \\
\frac{\partial l}{\partial \beta_{p-1}} \\[6pt]
\frac{\partial l}{\partial \ln(\sigma_{1})} \\
\vdots \\ 
\frac{\partial l}{\partial \ln(\sigma_{J})}
\end{pmatrix}, \ \text{for} \ p\ge2 \ \text{and} \ J\ge1.
\end{equation*}
Note that the case where $J=1$ is equivalent to the case in Appendix~\ref{grad_simple}. 

Taking the derivative of the log-likelihood with respect to $\beta_{k}$,
\begin{equation}\label{eqn:d_betak_sepvar}
\begin{split}
\frac{\partial l}{\partial \beta_{k}}=&\sum_{j=1}^{J}\Bigg\{\sum_{i=1}^{n_{j}}-\frac{x_{ijk}\phi(a_{ij}^{*})}{\sigma_{j}[1-\Phi(a_{ij}^{*})]} -\sum_{i\in S_{0j}}\frac{x_{ijk}\big[\phi(\nu_{ij}^{*})-\phi(a_{ij}^{*})\big]}{\sigma_{j}[\Phi(\nu_{ij}^{*})-\Phi(a_{ij}^{*})]} \\
&+\frac{1}{\sigma_{j}^{2}}\sum_{i\in S_{1j}}x_{ijk}(y_{ij}-\mathbf{x}_{ij}^{T}\boldsymbol{\beta})\Bigg\} \ \ \text{for} \ k = 1,\dots, p-1,
\end{split}
\end{equation}
and with respect to $\ln(\sigma_{j})$,
\begin{equation}\label{eqn:d_logsigma_sepvar}
\begin{split}
\frac{\partial l}{\partial \ln(\sigma_{j})}=&\sum_{i=1}^{n_{j}}-\frac{a_{ij}^{*}\phi(a_{ij}^{*})}{1-\Phi(a_{ij}^{*})}
-\sum_{i\in S_{0j}}\frac{\big[\nu_{ij}^{*}\phi(\nu_{ij}^{*})-a_{ij}^{*}\phi(a_{ij}^{*})\big]}{\Phi(\nu_{ij}^{*})-\Phi(a_{ij}^{*})} -n_{1j}\\
&+\frac{1}{\sigma_{j}^{2}}\sum_{i\in S_{1j}}(y_{ij}-\mathbf{x}_{ij}^{T}\boldsymbol{\beta})^{2} \ \ \text{for} \ j = 1,\dots, J.
\end{split}
\end{equation}

Therefore, the gradient vector is
\begin{equation}\label{eqn:gradient_sepvar}
\nabla l(\boldsymbol{\theta})=\begin{pmatrix}
\sum_{j=1}^{J}\sum_{i=1}^{n_{j}}-\frac{x_{ij1}\phi(a_{ij}^{*})}{\sigma_{j}[1-\Phi(a_{ij}^{*})]}
-\sum_{i\in S_{0j}}\frac{x_{ij1}\big[\phi(\nu_{ij}^{*})-\phi(a_{ij}^{*})\big]}{\sigma_{j}[\Phi(\nu_{ij}^{*})-\Phi(a_{ij}^{*})]}
+\frac{1}{\sigma_{j}^{2}}\sum_{i\in S_{1j}}x_{ij1}(y_{ij}-\mathbf{x}_{ij}^{T}\boldsymbol{\beta}) \\[6pt] \vdots \\[6pt]
\sum_{j=1}^{J}\sum_{i=1}^{n_{j}}-\frac{x_{ij(p-1)}\phi(a_{ij}^{*})}{\sigma_{j}[1-\Phi(a_{ij}^{*})]}
-\sum_{i\in S_{0j}}\frac{x_{ij(p-1)}\big[\phi(\nu_{ij}^{*})-\phi(a_{ij}^{*})\big]}{\sigma_{j}[\Phi(\nu_{ij}^{*})-\Phi(a_{ij}^{*})]}
+\frac{1}{\sigma_{j}^{2}}\sum_{i\in S_{1j}}x_{ij(p-1)}(y_{ij}-\mathbf{x}_{ij}^{T}\boldsymbol{\beta})\\[6pt]
\sum_{i=1}^{n_{1}}-\frac{a_{i1}^{*}\phi(a_{i1}^{*})}{1-\Phi(a_{i1}^{*})}
-\sum_{i\in S_{01}}\frac{\big[\nu_{i1}^{*}\phi(\nu_{i1}^{*})-a_{i1}^{*}\phi(a_{i1}^{*})\big]}{\Phi(\nu_{i1}^{*})-\Phi(a_{i1}^{*})} -n_{11}
+\frac{1}{\sigma_{1}^{2}}\sum_{i\in S_{11}}(y_{i1}-\mathbf{x}_{i1}^{T}\boldsymbol{\beta})^{2}\\[6pt] \vdots \\[6pt]
\sum_{i=1}^{n_{J}}-\frac{a_{iJ}^{*}\phi(a_{iJ}^{*})}{1-\Phi(a_{iJ}^{*})}
-\sum_{i\in S_{0J}}\frac{\big[\nu_{iJ}^{*}\phi(\nu_{iJ}^{*})-a_{iJ}^{*}\phi(a_{iJ}^{*})\big]}{\Phi(\nu_{iJ}^{*})-\Phi(a_{iJ}^{*})} -n_{1J}
+\frac{1}{\sigma_{J}^{2}}\sum_{i\in S_{1J}}(y_{iJ}-\mathbf{x}_{iJ}^{T}\boldsymbol{\beta})^{2}
\end{pmatrix}.
\end{equation}

\section{Hessian for Heteroskedastic Model}\label{hess_sepvar}
The Hessian matrix for parameters $\boldsymbol{\theta}$ in Appendix~\ref{grad_sepvar} is derived by taking further partial derivatives. This matrix takes the form
\begin{equation*}
\nabla^{2}l(\boldsymbol{\theta})=\begin{bmatrix}
\frac{\partial^{2} l}{\partial \beta_{1}^{2}} & \cdots & \frac{\partial^{2} l}{\partial \beta_{1} \partial\beta_{p-1}} & \frac{\partial^{2} l}{\partial \beta_{1} \partial\ln(\sigma_{1})} & \cdots & \frac{\partial^{2} l}{\partial \beta_{1} \partial\ln(\sigma_{J})}\\[6pt]
  \vdots & \ddots  & \vdots & \vdots  & \ddots  & \vdots \\[6pt]
  \frac{\partial^{2} l}{\partial\beta_{p-1} \partial \beta_{1}}& \cdots & \frac{\partial^{2} l}{\partial \beta_{p-1}^{2}} &  \frac{\partial^{2} l}{\partial \beta_{p-1} \partial \ln(\sigma_{1})} & \cdots & \frac{\partial^{2} l}{\partial \beta_{p-1} \partial \ln(\sigma_{J})}\\[6pt]
  \frac{\partial^{2} l}{\partial\ln(\sigma_{1}) \partial \beta_{1}} & \cdots & \frac{\partial^{2} l}{\partial \ln(\sigma_{1}) \partial \beta_{p-1}} &  \frac{\partial^{2} l}{\partial \ln^{2}(\sigma_{1})} & \cdots & \frac{\partial^{2} l}{\partial \ln(\sigma_{1})\partial \ln(\sigma_{J})} \\[6pt]
\vdots & \ddots & \vdots &  \vdots & \ddots & \vdots \\[6pt]
  \frac{\partial^{2} l}{\partial\ln(\sigma_{J}) \partial \beta_{1}} & \cdots & \frac{\partial^{2} l}{\partial \ln(\sigma_{J}) \partial \beta_{p-1}} &  \frac{\partial^{2} l}{\partial \ln(\sigma_{J})\partial \ln(\sigma_{1})} & \cdots &\frac{\partial^{2} l}{\partial \ln^{2}(\sigma_{j})}
\end{bmatrix}.
\end{equation*}
Note that since the groups are assumed to be independent, $\frac{\partial^{2} l}{\partial \ln(\sigma_{j})\partial \ln(\sigma_{k})}=0$ for all $j\ne k$, which reduces the Hessian matrix to
\begin{equation*}
\nabla^{2}l(\boldsymbol{\theta})=\begin{bmatrix}
\frac{\partial^{2} l}{\partial \beta_{1}^{2}} & \cdots & \frac{\partial^{2} l}{\partial \beta_{1} \partial\beta_{p-1}} & \frac{\partial^{2} l}{\partial \beta_{1} \partial\ln(\sigma_{1})} & \cdots & \frac{\partial^{2} l}{\partial \beta_{1} \partial\ln(\sigma_{J})}\\[6pt]
  \vdots & \ddots  & \vdots & \vdots  & \ddots  & \vdots \\[6pt]
  \frac{\partial^{2} l}{\partial\beta_{p-1} \partial \beta_{1}}& \cdots & \frac{\partial^{2} l}{\partial \beta_{p-1}^{2}} &  \frac{\partial^{2} l}{\partial \beta_{p-1} \partial \ln(\sigma_{1})} & \cdots & \frac{\partial^{2} l}{\partial \beta_{p-1} \partial \ln(\sigma_{J})}\\[6pt]
  \frac{\partial^{2} l}{\partial\ln(\sigma_{1}) \partial \beta_{1}} & \cdots & \frac{\partial^{2} l}{\partial \ln(\sigma_{1}) \partial \beta_{p-1}} &  \frac{\partial^{2} l}{\partial \ln^{2}(\sigma_{1})} &  & 0 \\[6pt]
\vdots & \ddots & \vdots &   & \ddots &  \\[6pt]
  \frac{\partial^{2} l}{\partial\ln(\sigma_{J}) \partial \beta_{1}} & \cdots & \frac{\partial^{2} l}{\partial \ln(\sigma_{J}) \partial \beta_{p-1}} &  0 &  &\frac{\partial^{2} l}{\partial \ln^{2}(\sigma_{j})}
\end{bmatrix}.
\end{equation*}

The individual components of this matrix are calculated as
\begin{equation}\label{eqn:d_betak_betal_sepvar}
\scriptsize
\begin{split}
\frac{\partial^{2} l}{\partial \beta_{k}\partial\beta_{l}}=&
\sum_{j=1}^{J}\Bigg\{\sum_{i=1}^{n_{j}}-\frac{x_{ijk}x_{ijl}\Big\{a_{ij}^{*}[1-\Phi(a_{ij}^{*})]\phi(a_{ij}^{*})-\phi^{2}(a_{ij}^{*})\Big\}}{\sigma_{k}\sigma_{l}[1-\Phi(a_{ij}^{*})]^{2}}\\
&-\sum_{i\in S_{0j}}\frac{x_{ijk}x_{ijl}\Big\{[\nu_{ij}^{*}\phi(\nu_{ij}^{*})-a_{ij}^{*}\phi(a_{ij}^{*})][\Phi(\nu_{ij}^{*})-\Phi(a_{ij}^{*})]+[\phi(\nu_{ij}^{*})-\phi(a_{ij}^{*})]^{2}\Big\}}{\sigma_{k}\sigma_{l}[\Phi(\nu_{ij}^{*})-\Phi(a_{ij}^{*})]^{2}}\\
&-\frac{1}{\sigma^{2}_{j}}\sum_{i\in S_{1j}}x_{ijk}x_{ijl}\Bigg\}\ \text{for} \ k= 1,\dots,p-1 \ \text{and} \ l = k,\dots,p-1,
\end{split}
\end{equation}

\begin{equation}\label{eqn:d_betak_logsigmaj_sepvar}
\scriptsize
\begin{split}
\frac{\partial^{2} l}{\partial \beta_{k} \partial \ln(\sigma_{j})}=&
\sum_{i=1}^{n_{j}}\frac{x_{ijk}\Big\{[1-\Phi(a_{ij}^{*})]\phi(a_{ij}^{*})[1-(a_{ij}^{*})^{2}]+a_{ij}^{*}\phi^{2}(a_{ij}^{*})\Big\}}{\sigma_{j}[1-\Phi(a_{ij}^{*})]^{2}}\\
&-\sum_{i\in S_{0j}} \frac{x_{ijk}\bigg\{\Big(\Phi(\nu_{ij}^{*})-\Phi(a_{ij}^{*})\Big)\Big(\phi(\nu_{ij}^{*})\big[1-(\nu_{ij}^{*})^{2}\big]-\phi(a_{ij}^{*})\big[1-(a_{ij}^{*})^{2}\big]\Big)-\Big[\phi(\nu_{ij}^{*})-\phi(a_{ij}^{*})\Big]\Big[\phi(\nu_{ij}^{*})\nu_{ij}^{*}-\phi(a_{ij}^{*})a_{ij}^{*}\Big]\bigg\}}{\sigma_{j}[\Phi(\nu_{ij}^{*})-\Phi(a_{ij}^{*})]^{2}}\\
&-\frac{2}{\sigma_{j}^{2}}\sum_{i\in S_{1j}}x_{ijk}(y_{ij}-\mathbf{x}_{ij}^{T}\boldsymbol{\beta}) \ \text{for} \ k = 1,\dots,p-1 \ \text{and} \ j = 1,\dots,J,
\end{split}
\end{equation}

\begin{equation}\label{eqn:d_logsigmaj_2_sepvar}
\scriptsize
\begin{split}
\frac{\partial^{2} l}{\partial \ln^{2}(\sigma_{j})}=&
\sum_{i=1}^{n_{j}}\frac{a_{ij}^{*}\Big\{[1-\Phi(a_{ij}^{*})]\phi(a_{ij}^{*})[1-(a_{ij}^{*})^{2}]+a_{ij}^{*}\phi^{2}(a_{ij}^{*})\Big\}}{[1-\Phi(a_{ij}^{*})]^{2}}\\
&-\sum_{i\in S_{0j}}\frac{\Big(\Phi(\nu_{ij}^{*})-\Phi(a_{ij}^{*})\Big)\Big(\phi(\nu_{ij}^{*})\big[(\nu_{ij}^{*})^{3}-\nu_{ij}^{*}\big]-\phi(a_{ij}^{*})\big[(a_{ij}^{*})^{3}-a_{ij}^{*}\big]\Big)+\Big[\phi(\nu_{ij}^{*})\nu_{ij}^{*}-\phi(a_{ij}^{*})a_{ij}^{*}\Big]^{2}}
{[\Phi(\nu_{ij}^{*})-\Phi(a_{ij}^{*})]^{2}}\\
&-\frac{2}{\sigma_{j}^{2}}\sum_{i\in S_{1j}}(y_{ij}-\mathbf{x}_{ij}^{T}\boldsymbol{\beta})^{2} \ \text{for} \ j = 1,\dots, J.
\end{split}
\end{equation}



\clearpage

%

\end{document}